\documentclass[12pt,preprint]{aastex}

\usepackage{ulem}
\usepackage{psfig}

\def\gtorder{\mathrel{\raise.3ex\hbox{$>$}\mkern-14mu
             \lower0.6ex\hbox{$\sim$}}}
\def\ltorder{\mathrel{\raise.3ex\hbox{$<$}\mkern-14mu
             \lower0.6ex\hbox{$\sim$}}}

\newcommand{\Swift}{\textit{Swift}}
\newcommand{\event}{\textit{Swift} J195509.6+261406}

\slugcomment{ \today}
\shorttitle{Galactic GRB070610}
\shortauthors{Kasliwal, Cenko, Kulkarni et al.}
\begin{document}

\title{GRB\,070610: A Curious Galactic Transient}
\author{
M.~M.~Kasliwal\altaffilmark{1},
S.~B.~Cenko\altaffilmark{2},
S.~R.~Kulkarni\altaffilmark{1},
P.~B.~Cameron\altaffilmark{1},
E.~Nakar\altaffilmark{1},
E.~O.~Ofek\altaffilmark{1},
A.~Rau\altaffilmark{1},
A.~M.~Soderberg\altaffilmark{1},
S.~Campana\altaffilmark{3},
J.~S.~Bloom\altaffilmark{4},
D.~A.~Perley\altaffilmark{4},
L.~K.~Pollack\altaffilmark{17},
S.~Barthelmy\altaffilmark{5},
J.~Cummings\altaffilmark{5},
N.~Gehrels\altaffilmark{5},
H.~A.~Krimm\altaffilmark{6,16},
C.~B.~Markwardt\altaffilmark{6,7},
G.~Sato\altaffilmark{5},
P.~Chandra\altaffilmark{8},
D.~Frail\altaffilmark{9},
D.~B.~Fox\altaffilmark{10},
P.~Price\altaffilmark{11}, 
E.~Berger\altaffilmark{12,13},
S.~A.~Grebenev\altaffilmark{14}, 
R.~A.~Krivonos\altaffilmark{14,15} \& \
R.~A.~Sunyaev\altaffilmark{14,15}
}
\altaffiltext{1}{Division of Physics, Mathematics and Astronomy, 
California Institute of Technology, MS 105-24, Pasadena, CA 91125, USA}
\altaffiltext{2}{Space Radiation Laboratory, California Institute of 
Technology, MS 220-47, Pasadena, CA 91125, USA}
\altaffiltext{3}{INAF, Osservatorio Astronomica di Brera,
via E. Bianchi 46, I-23807 Merate (LC), Italy}
\altaffiltext{4}{Department of Astronomy, University of California, Berkeley, CA 94720, USA}
\altaffiltext{5}{NASA Goddard Space Flight Center, Greenbelt, MD~20771, USA}
\altaffiltext{6}{CRESST and Astroparticle Physics Laboratory, NASA/GSFC, 
Greenbelt, MD 20771, USA}
\altaffiltext{7}{Department of Astronomy, University of Maryland, 
College Park, MD 20742, USA}
\altaffiltext{8}{University of Virginia, P.O. Box 400325, Charlottesville, 
VA 22903, USA}
\altaffiltext{9}{National Radio Astronomy Observatory, Socorro, NM~87801, USA}
\altaffiltext{10}{Department of Astronomy \&\ Astrophysics, 525 Davey
Laboratory, Pennsylvania State University, University Park, PA~16802, USA}
\altaffiltext{11}{Institute for Astronomy, University of Hawaii,
2680 Woodlawn Drive, Honolulu, HI~96822, USA}
\altaffiltext{12}{Observatories of the Carnegie Institute of Washington,
Pasadena, CA 91101, USA}
\altaffiltext{13}{Princeton University Observatory, Princeton, NJ 08544, USA}
\altaffiltext{14}{Space Research Institute, Profsoyuznaya 84/32, 117997 
Moscow, Russia}
\altaffiltext{15}{Max-Plank-Institut fuer Astrophysik, Karl-Schwarzschild-Str. 
1, D-85741 Garching, Germany}
\altaffiltext{16}{Universities Space Research Association, 10211 Wincopin Circle, Suite 500, Columbia, MD 21044}
\altaffiltext{17}{Department of Astronomy and Astrophysics, University of California, Santa Cruz, CA 95064}

\begin{abstract}
GRB\,070610 is a typical high-energy event with a duration of 5\,s.
Yet within the burst localization we detect a highly unusual X-ray
and optical transient, \event.  We see high amplitude X-ray and optical
variability on very short time scales even at late times.
Using near-infrared imaging assisted by a laser guide star and
adaptive optics, we identified the counterpart of \event.
Late-time optical and near-infrared imaging constrain the spectral type 
of the counterpart to be fainter than a K-dwarf assuming it is of 
Galactic origin. It is possible that GRB\,070610 
and \event\ are unrelated sources. However, the absence of a typical X-ray 
afterglow from GRB\,070610 in conjunction 
with the spatial and temporal coincidence of the two motivate 
us to suggest that the sources are related. The closest (imperfect) 
analog to \event\ is V4641~Sgr, an unusual black hole binary. 
We suggest that \event\ along with V4641~Sgr 
define a sub-class of stellar black hole binaries -- the fast X-ray novae.
We further suggest that fast X-ray novae are associated with
bursts of gamma-rays. If so, GRB\,070610 defines a new class of
celestial gamma-ray bursts and these bursts dominate the long-duration 
GRB demographics.
\end{abstract}


\facility{{\it Facilities:} \facility{PO:1.5m}, \facility{Hale}, \facility{Keck:I}, \facility{Keck:II}, \facility{VLA}, \facility{Swift}}

\section{Discovery of GRB\,070610}
\label{sec:BAT}

Launched in November 2004, the 
\Swift\ Gamma-Ray Burst Explorer \citep{gcg+04} 
was designed to localize $\gamma$-ray bursts (GRBs) and undertake
rapid and sustained X-ray and Ultra-Violet observations of the resulting afterglow.
With over two hundred events now localized and studied, \Swift\ has
made fundamental contributions to both long-duration soft
bursts (LSBs) and short-duration hard bursts (SHBs). LSBs appear to
trace cosmological massive-star formation rate with one event at a redshift
of 6.3. SHBs have been seen at typical redshifts of $\sim 0.5$ in
both elliptical and star-forming galaxies. There is now some 
circumstantial evidence for SHBs being the result of coalescence of 
compact objects.

At 20:52:26 UT on 2007 June 10 the Burst Alert Telescope (BAT; 
\citealt{bbc+05})
aboard \Swift\ triggered on GRB\,070610.  The high-energy prompt emission
had a duration ($T_{90}$) of 4.6\,s \citep{Pagani07_GCN6489}.
Over the range 15--150\,keV the burst could be fitted with a
power law with photon index $\Gamma =1.76\pm 0.25$, resulting in a fluence of
$(2.4\pm 0.4)\times 10^{-7}\,{\rm erg\,cm^{-2}}$
\citep{Tueller07_GCN6491}. A blackbody model is inconsistent with this
emission (reduced $\chi^2 = $1.7). 

\begin{figure}[htb]
\centerline{\psfig{file=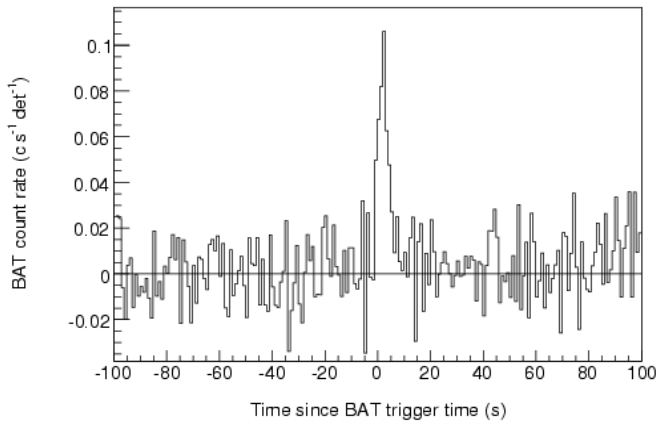,width=10cm,angle=0}}
\caption[]{\small
15-150 keV \Swift-BAT light curve of GRB\,070610, with 1-s time resolution.
The conversion factor to translate the ordinate to cgs flux units is
$5.6 \times 10^{-7}$\,erg cm$^{-2}$ c$^{-1}$ det (1 det = 0.16 cm$^{2}$).
}
\label{fig:BATProfile}
\end{figure}

\begin{figure}[htb]
\centerline{\psfig{file=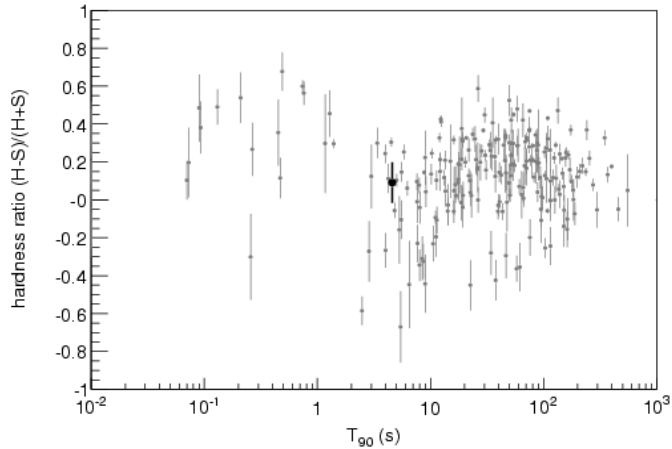,width=10cm,angle=0}}
\caption[]{\small
Plot of duration ($T_{90}$) and hardness ratio (HR) of 226 \Swift\
bursts from GRB\,041217 to GRB\,070616.  We define hardness ratio as
$(H-S)/(H+S)$, where $S$ and $H$ are energy fluences in 15--50 keV and
50--150 keV, respectively.  
The values of $T_{90}$ and hardness ratio for GRB\,070610
(marked by a large filled black circle) are 
$4.6\pm 0.4\,$s and 0.09 $\pm$ 0.11 (90\% confidence level),
respectively.
}
\label{fig:T90-HR}
\end{figure}

The burst profile consisted of a single symmetric peak (Figure 
\ref{fig:BATProfile}).  Fitting the profile \citep{nnb+96}, we calculate
a rise time (i.e.~half width at half maximum) of $1.68 \pm 0.55$\,s. 
As can be seen from Figure~\ref{fig:T90-HR}, the duration 
and the hardness ratio of \event\ are both consistent with the
broader population of extragalactic long-duration GRBs observed by \Swift.

\begin{figure}[htb]
\centerline{\psfig{file=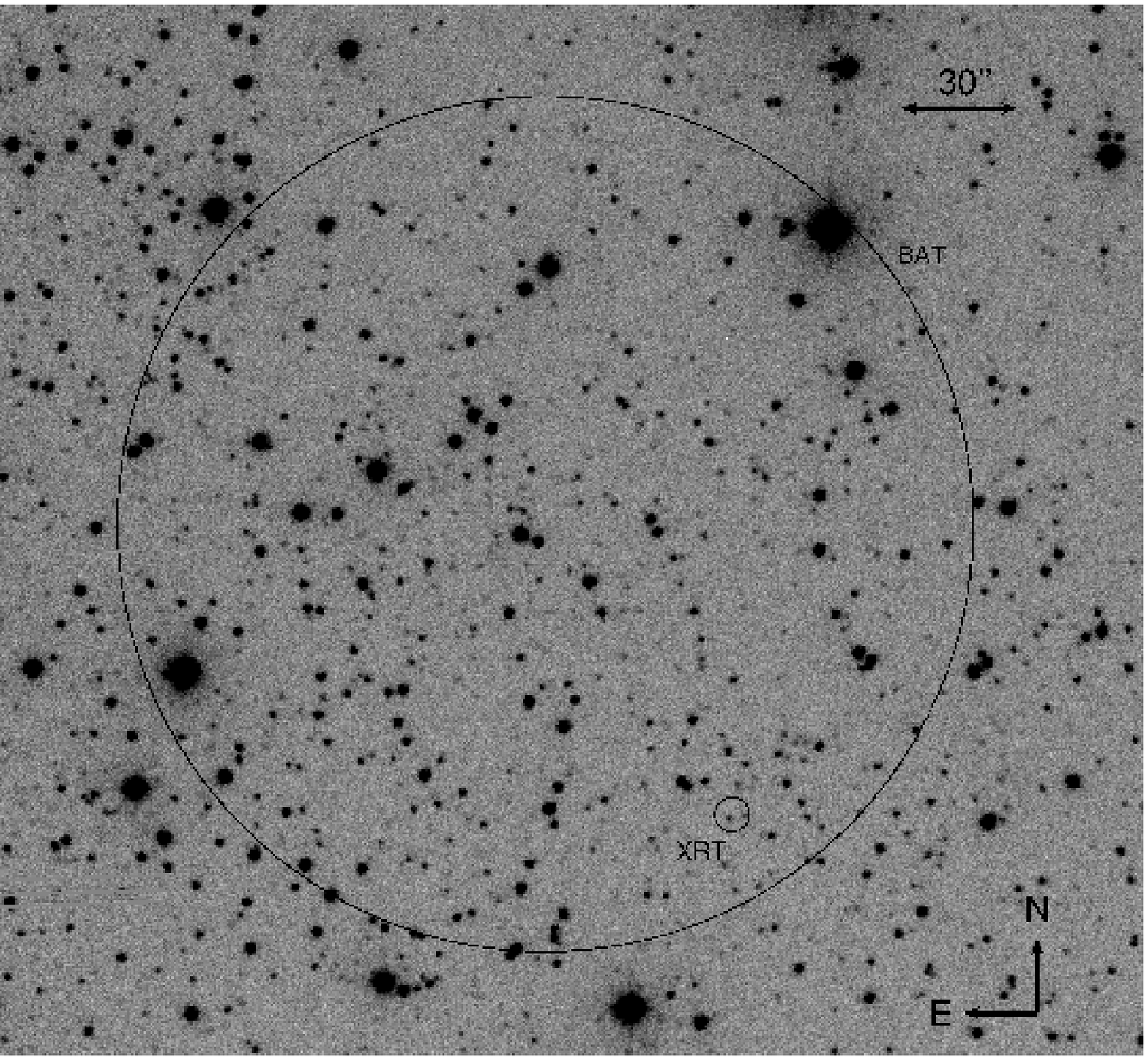,width=8.5cm}}
\caption[]{\small
Optical image ($i^{\prime}$-band) of the field of GRB\,070610 obtained by the 
automated Palomar 60-inch telescope on UT 2007 June 12.  
The BAT localization of GRB\,070610 has a radius of 1.8\arcmin, while 
the XRT localization of \event\ has a radius of 4.3\arcsec; both are indicated
with black circles. The bright source in the XRT circle is
\event\ .  
}
\label{fig:P60image}
\end{figure}

The BAT localized GRB\,070610 to $\alpha=19^{\mathrm{h}} 55^{\mathrm{m}}
13\farcs1$, $\delta=+26^{\circ} 15\arcmin 20\arcsec$ (J2000.0)
and a 90\%-containment radius of 1.8\arcmin.  As
can be seen in Figure~\ref{fig:P60image} the field is dense, which
is not surprising given the Galactic location ($l=63.3^\circ$ and
$b=-1.0^\circ$).

Here we report the discovery of an unusual X-ray transient
(hereafter referred to as \event) in the error circle of GRB\,070610
and followup optical, near-infrared (NIR) and radio observations.

\section{\event: A Transient X-ray Source}
\label{sec:XRT}

The X-ray Telescope (XRT; \citealt{bhn+05}) began observing the field of 
GRB\,070610 3.2\,ks after the initial BAT trigger (prompt slewing was disabled
due to an Earth limb constraint).  The XRT detected a single uncatalogued 
variable source in the BAT error circle at $\alpha=19^{\mathrm{h}} 55^{\mathrm{m}} 
9\farcs6$, $\delta=+26^{\circ} 14\arcmin 6\farcs7$ (90\%
confidence error circle of $4\farcs3$\ radius; \citealt{Pagani07_GCN6506}).
This position was further refined to  $\alpha=19^{\mathrm{h}} 55^{\mathrm{m}} 
9\farcs66$, $\delta=+26^{\circ} 14\arcmin 5\farcs2$ (90\%
confidence error circle of 1$\farcs2\ $radius~\footnote{http://astro.berkeley.edu/$\sim$nat/swift/xrt\_pos.html}).

The XRT continued to monitor \event\ over the course of the next month
until the source was no longer detected.

The XRT data were processed with \texttt{xrtpipeline
(v0.10.6)}.  All data were obtained in photon counting mode. In this mode the 
entire CCD is read and the time resolution is
limited to 2.5\,s.  We extracted grade 0--12 events \citep{bhn+05} from a 15 
pixel radius circular region centered on the source. To account for the 
background, we extracted events within a 40 pixel radius circular region in
the vicinity of the transient but not encompassing any other source
in the field.
We adaptively extracted the light curve binning the data in order
to have 10 counts per bin. The light curve was corrected for the extraction
region losses and for CCD defects as well as for vignetting by using
the task \texttt{xrtlccorr (v0.1.9)}, which generates an orbit-by-orbit
correction based on the instrument map.

\begin{figure*}[htb]
\centerline{\psfig{file=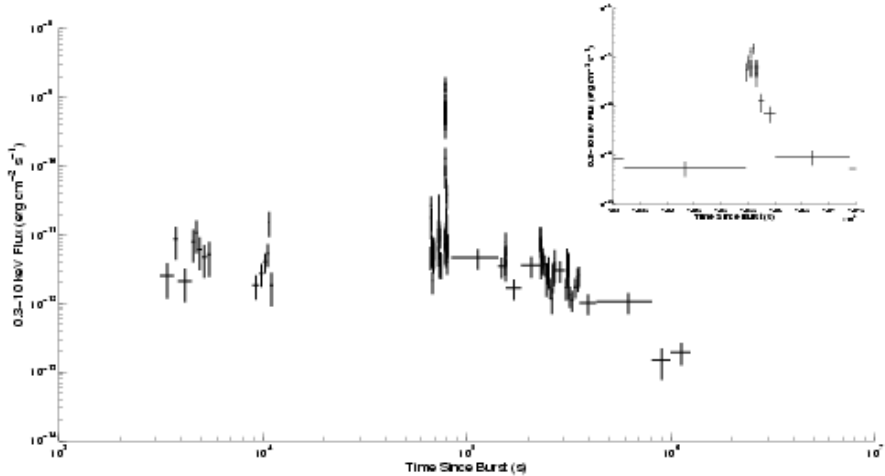,width=14cm,angle=0}}
\caption[]{\small
XRT light curve of \event\ in the energy band 0.3--10\,keV.
The dramatic X-ray flare
at $t \sim 7.86 \times 10^{4}$\, s is shown in the inset.}
\label{fig:XRT}
\end{figure*}

\begin{figure}[hbt]
\centerline{\psfig{file=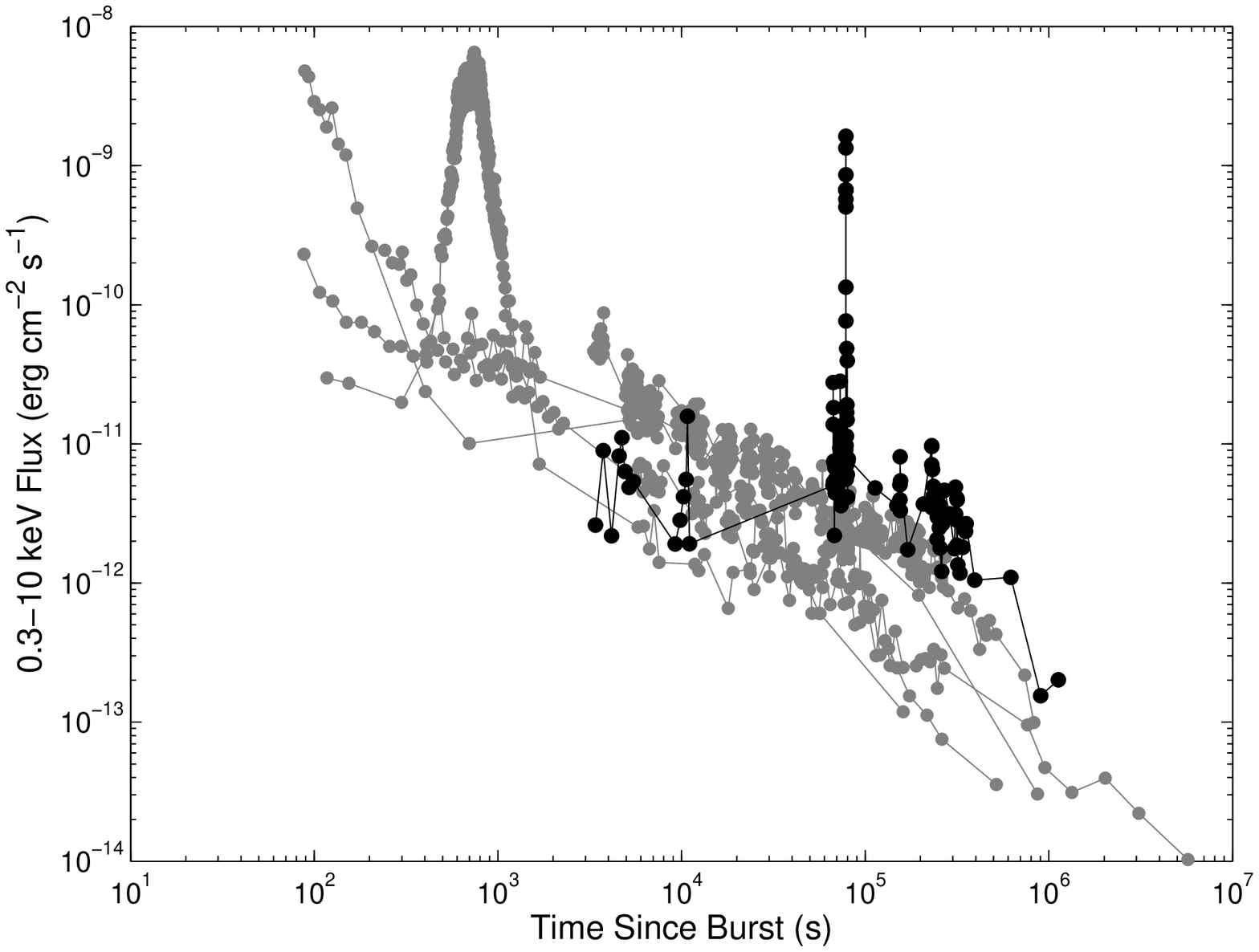,width=10cm,angle=0}}
\caption[]{\small
XRT light curves of a small sample of extragalactic long-duration
GRBs (GRBs\,050315, 050318, 050319, 050416A, and 050502B) are shown in
grey.  Data are from \citealt{ebp+07}. All show the
approximately power-law decay typical of GRB afterglows. GRB\,050502B
exhibits a bright flare around $t\sim 10^3\,$s (see \citealt{fbl+06}).
However, the rise time of this flare is much longer than the
spike seen in \event\ (shown in black). 
}
\label{fig:xrtmoal}
\end{figure}

The X-ray light curve of \event\ is shown in Figure \ref{fig:XRT} and compared
to a small sample of long-duration GRB afterglows in Figure \ref{fig:xrtmoal}.
\citealt{Kann07_GCN6505} were the first to suggest that this GRB was likely to be of Galactic origin.
Clearly, \event\ differs from typical GRB X-ray afterglows in two fundamental respects.  
First, it does not exhibit the strong (overall) secular decrease
in flux over timescales of hours \citep{nkg+06,zfd+06}. While the 
decay index in long-duration GRBs can vary markedly from one phase to another, 
\event\ shows no significant decline until very late times ($\sim 10^{6}$\, s).

Secondly, the XRT light curve of \event\ consists of spikes -- never seen
before in any afterglow.   In particular we 
draw the attention of the reader to a dramatic flare at 
$t \sim 7.86 \times 10^{4}$\,s, jumping by a factor of $\Delta f / f \sim 100$ in 
flux over a time scale of $\Delta t / t \sim 10^{-4}$ 
(see Figure \ref{fig:XRT}, inset). None of the sixty nine XRT flares 
described in \citet{cmr+07} exhibit a comparable amplitude 
spike at late time. While a strong X-ray flare has been seen 
in GRB\,050502B \citep{fbl+06}
(see Figure~\ref{fig:xrtmoal}) the fractional duration,
$\Delta t/t$ is much larger ($\sim 0.5$). Less significant variability
is present throughout the duration of observations of \event\ .

We searched the XRT flare for pulsations. 521 photons were extracted
withing 60\arcsec of the source position and corrected to the solar
system barycenter with the task \texttt{barycorr}.  To search for
pulsations we constructed the $Z^2_1$ power spectrum to a maximum
frequency of 0.2\,Hz \citep{bbb+83}. The largest observed value of
$Z^2_1$ was 25.2 at a frequency of 0.1446\,Hz. Since $Z^2_n$ is
distributed as $\chi^2$ with $2n$ degrees of freedom, this value
corresponds to a single trial detection significance of 4.8$\sigma$ in
equivalent Gaussian units. Given that we have performed 350 trials,
the significance of this detection is 3.4$\sigma$, and thus we do not
consider this result to be conclusive evidence of peroidicity.

For spectral analysis the ancillary response files were generated
with the task \texttt{xrtmkarf}. We used the latest spectral
redistribution matrices (v009).  Data were extracted from single
or consecutive orbits in order to have at least 100 counts per
spectrum. Spectra were binned to a minimum of 15 counts per energy
bin. The resulting spectra were inconsistent with a blackbody
(reduced $\chi^2 = $1.9) and consistent with a power law model 
(task \texttt{phabs}).
The best fit column density ($N_{\mathrm{H}}$)
and photon index ($\Gamma$) for each epoch are summarized in Table
\ref{tab:XRTSpectrum}.  Overall, we find that the inferred flux conversion
is approximately 1 count s$^{-1} \approx 1.3 \times 10^{-10}$\,erg cm$^{-2}$
s$^{-1}$ in the 0.3--10\,keV band. 

We extrapolate the XRT flare spectrum to BAT (15--50 keV) and predict a
flux of 1.8 $\times$ 10$^{-9}$\,erg cm$^{-2}$s$^{-1}$. This corresponds to
a BAT count rate of 0.0032 counts~s $^{-1}$~det$^{-1}$. This is consistent
with a 2-$\sigma$ upper limit from two 64s intervals of BAT data straddling 
the XRT flare --- 0.0038 counts~s $^{-1}$~det$^{-1}$ (at 78499.8\,s) and 
0.012 counts~s $^{-1}$~det$^{-1}$(at 78563.8\,s).

The inferred interstellar extinction along this low Galactic latitude
is quite high and thus uncertain: $N_{\rm H}$ of $1.1\times
10^{22}\,$cm$^{-2}$ \citep{dl90}; $0.8\times 10^{22}\,$cm$^{-2}$
\citep{kbh+05}; and 1.56-1.89$\times 10^{22}\,$cm$^{-2}$ \citep{sfd98}.
The former two estimates are based on H~I data whereas the latter
on diffuse infrared emission.  Given the uncertainty in the inferred
$N_{\mathrm{H}}$ the XRT spectrum cannot be used to determine the distance to
\event.

\begin{deluxetable}{llll}
  \tabletypesize{\footnotesize}
  \tablecaption{XRT Spectral Analysis}
  \tablecolumns{4}
  \tablewidth{0pc}
  \tablehead{\colhead{Epoch Start} & \colhead{Total Exposure} & 
             \colhead{$N_{\mathrm{H}}$} & \colhead{$\Gamma$} \\
             \colhead{(MJD)} & \colhead{(s)} & \colhead{($10^{22}$ cm$^{-2}$)} &  }
  \startdata
    54261.907 & 4811 & $0.30^{+0.29}_{-0.23}$ & 1.43 $\pm$ 0.37  \\
    54262.641 & 7912 & $0.76^{+0.24}_{-0.18}$ & 1.93 $\pm$ 0.18  \\
    54263.268, 54264.004 & 2947, 10500 & $0.59^{+0.31}_{-0.23}$ & 
     1.11 $\pm$ 0.22  \\
    54265.387 & 6026 & $0.61^{+0.53}_{-0.33}$ & 1.33 $\pm$ 0.40  \\
    Flare    & \ldots & $0.92^{+0.91}_{-0.57}$ & 1.74 $\pm$ 0.48 \\
    All but flare & \ldots & $0.72^{+0.14}_{-0.12}$ & 1.71 $\pm$ 0.11 \\
  \enddata
  \tablecomments{We have fit the XRT data to a power-law model of the form
    $N(E) \propto E^{-\Gamma}$, leaving the line-of-sight $N_{\mathrm{H}}$
    as a free parameter.}
\label{tab:XRTSpectrum}
\end{deluxetable}

\section{A Flickering Optical Variable}
\label{sec:Optical}

\begin{figure}[hbt]
\centerline{\psfig{file=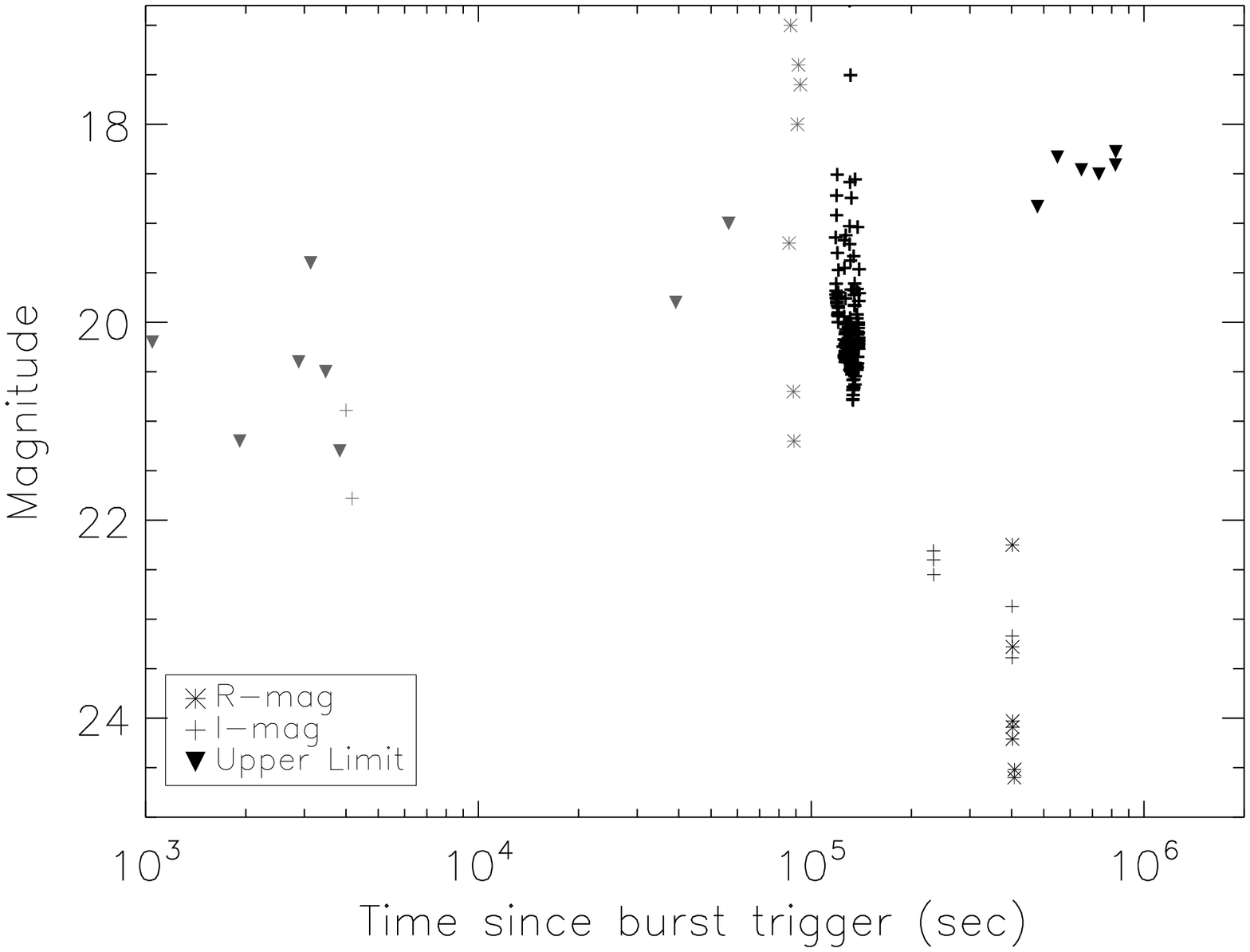,width=10cm,angle=90}}
\caption[]{\small 
Optical light curve of \event, including data
from P60 (black), Keck/LRIS (black), and the literature (grey).
\citep{French07_GCN6500,Postigo07_GCN6501,Kann07_GCN6505,GCN.6507,GCN.6508,GCN.6512,Klotz07_GCN6513, GCN.6532,GCN.6536}.
}
\label{fig:fullLC}
\end{figure}

\begin{figure*}[hbt]
\centerline{\psfig{file=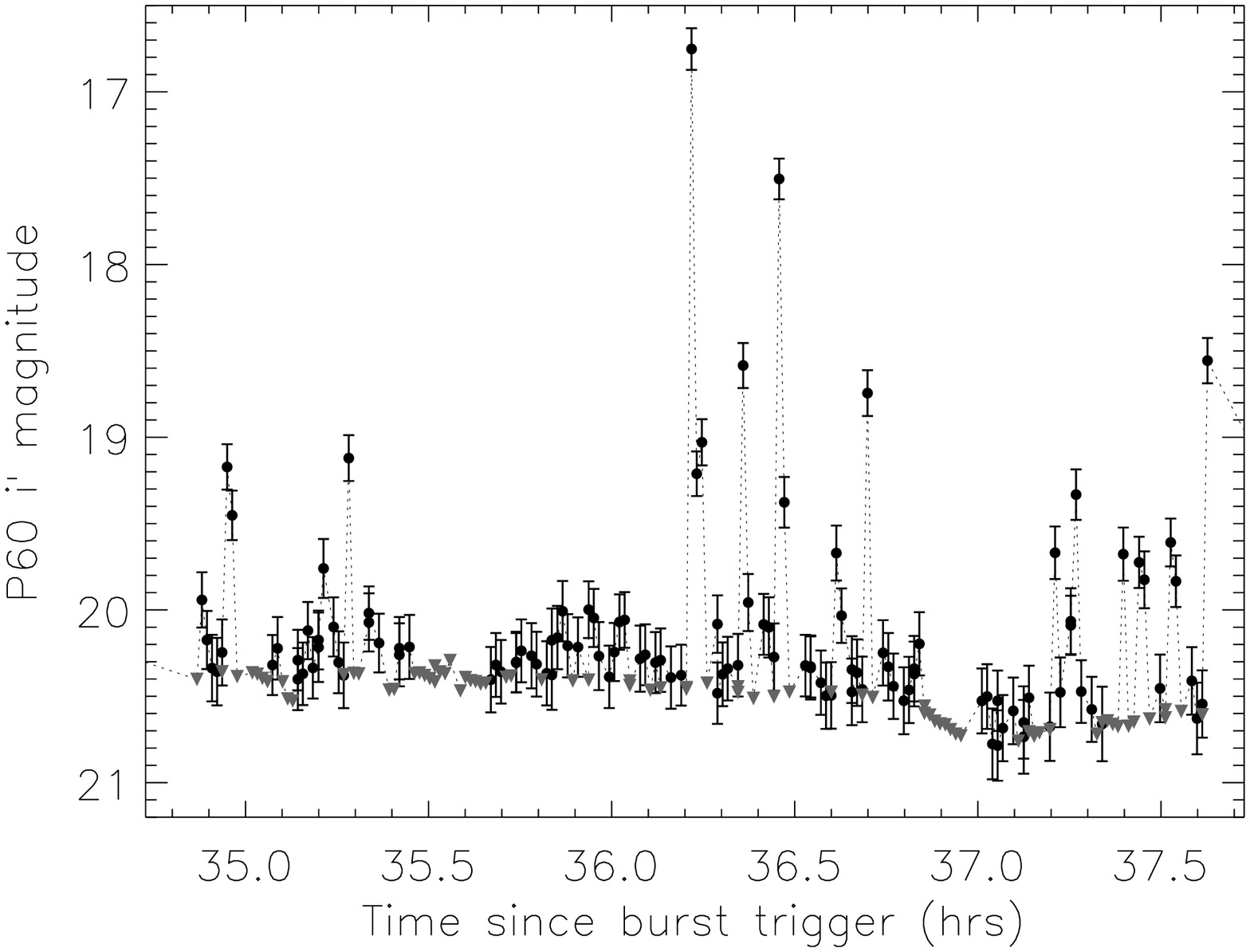,width=13cm,angle=90}}
\caption[]{\small 
P60 $i^{\prime}$-band light curve from the night of 2007 June 12.  
Upper limits are indicated by grey inverted triangles. The rapid
variability (time scales less than 60\,s, our sampling rate) at late 
times is unlike any previous long-duration GRB optical afterglow.
}
\label{fig:P60LC}
\end{figure*}

Rapid observations in response to the BAT trigger, in particular by the
\textit{OPTIMA-Burst} team \citep{Stefanescu07_GCN6492}, revealed a rapidly variable
(time scales as low as tens of seconds) optical transient inside the 
XRT error circle of \event.  
Astronomers using other facilities -- 
including the OSN 1.5-m telescope
\citep{Postigo07_GCN6501}, the 2-m Schmidt telescope of the
Th\"uringer Landessternwarte \citep{Kann07_GCN6505}, the
25-cm \textit{TAROT} facility \citep{Klotz07_GCN6513}, and the 40-cm
\textit{Watcher} telescope \citep{French07_GCN6500} -- confirmed
the detection of this variable source. Detections and upper limits reported to 
the GRB Coordinates Network (GCN\footnote{http://gcn.gsfc.nasa.gov/gcn3\_archive.html}) are shown in Figure \ref{fig:fullLC}.

Drawn by the excitement of these discoveries, we began monitoring
the field of \event\ in the $i^{\prime}$ filter with the automated 
Palomar 60-inch telescope (P60; \citealt{cfm+06}) starting at 5:47 
UT 2007 June 12 and continued over the next several nights.  
In addition, we imaged the field in $R$-, $I$- and $g$- bands with the 
{\it Low Resolution Imaging Spectrograph} (LRIS; \citealt{occ+95}) mounted 
at the Cassegrain focus of the Keck~I 10-m telescope.  All images were
reduced using standard IRAF\footnote{IRAF is distributed
by the National Optical Astronomy Observatory, which is operated by the
Association for Research in Astronomy, Inc., under cooperative agreement
with the National Science Foundation.} routines.  

The light curve obtained from our observations is also 
summarized in Figure~\ref{fig:fullLC}. The P60 and the Keck photometry 
can be found in Table~\ref{tab:p60lc} and Table~\ref{tab:kecklc} respectively.

\begin{deluxetable}{llllll}
  \tabletypesize{\footnotesize}
  \tablecaption{Optical Observations of \event\ at Keck I and Palomar Hale}
  \tablecolumns{6}
  \tablewidth{0pc}
  \tablehead{\colhead{Mean Epoch} & \colhead{Facility} &
             \colhead{Filter}  &
             \colhead{Exposure} & \colhead{Magnitude} \\
             \colhead{(2007 UT)} & & & \colhead{(s)} & \colhead{(s)} &
            }
  \startdata
Jun 13.570 & LRIS & $I$  & 120 $\times$ 3 & $> 24.0$ \\
Jun 15.517 & LRIS & $I$  & 200 $\times$ 3 & $24.37 \pm 0.21$ \\
Jun 15.524 & LRIS & $R$  & 180 $\times$ 1 & $22.25 \pm 0.06$ \\
Jun 15.527 & LRIS & $R$  & 180 $\times$ 1 & $24.21 \pm 0.13$ \\ 
Jun 15.531 & LRIS & $R$  & 180 $\times$ 1 & $23.28 \pm 0.07$ \\
Jun 15.534 & LRIS & $R$  & 180 $\times$ 1 & $24.09 \pm 0.11$ \\
Jun 15.594 & LRIS & $R$  & 45 $\times$ 8 & $> 25.0$\\
Aug 13.336 & LRIS & $R$  & 300 $\times$ 4 & $> 26.0$\\
Sep 13.362 & LFC  & $i^{\prime}$  & 360 $\times$ 26 &  $> 24.5$\\

 \enddata
 \tablecomments{Zeropoints computed in the Vega system. Error quoted 
        are 1-$\sigma$ photometric and instrumental 
        errors summed in quadrature.  Upper limits quote are 3-$\sigma$.  No
        correction has been made for the large line-of-sight
        extinction.}
\label{tab:kecklc}
\end{deluxetable}

The P60 light curve is dominated by flickering and magnificent flares
on the night of UT 2007 June 12 (see Figure \ref{fig:P60LC}).  
We observed over eleven flares with amplitudes greater 
than one magnitude in only three hours. The brightest of these flares
rose and dropped by more than 3.5 magnitudes within 6 minutes.
The amplitude of the flares is a lower limit because the P60 
images are not deep enough to detect the quiescent counterpart
(see below). The timescale is also an upper limit because it is 
entirely possible that variability is more rapid than our 
sampling rate ($\sim 60$\,s). If we define duty-cycle as the fraction
of time for which the \event\ was brighter than $i^{\prime} < 20$, then the 
duty cycle based on the first night of data is 18.6$\%$. Given
that there was no detection on subsequent ten nights, the duty
cycle reduces to 5.8$\%$. 

\begin{figure*}[htb]
\centerline{\psfig{file=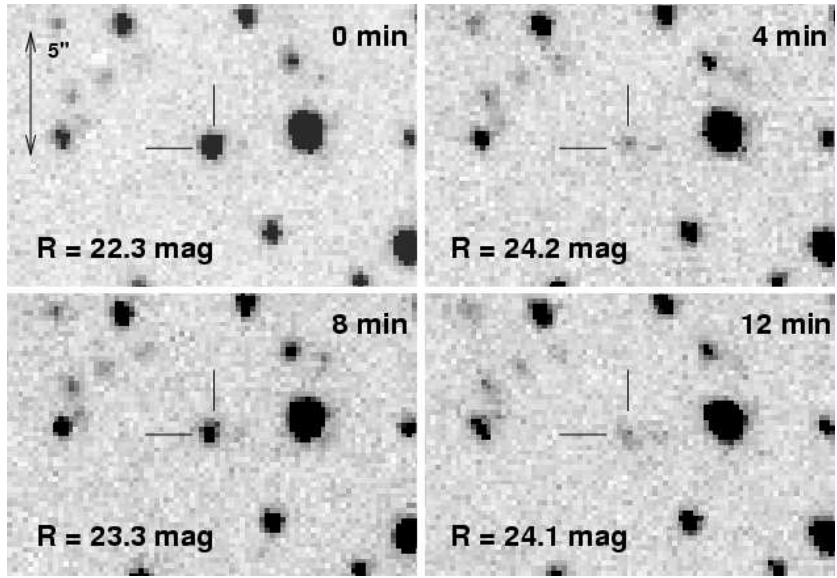,width=11cm}}
\caption[]{\small
Close up view of the optical field of the \event\ optical transient
using the \textit{LRIS} instrument on the Keck I 10-m telescope; 2007 June 15 
starting at 12:33 UT. All four images were taken in the $R$-band with 
a 180\,s exposure in sequence.  The transient brightens by over two magnitudes
in only three minutes about five days after the burst trigger.  Such
rapid variability at late times is unprecedented from an
extragalactic GRB optical afterglow.  
}
\label{fig:KeckFlare}
\end{figure*}

We see a dramatic flare in the \textit{LRIS} data five days 
(UT 2007 June 15) 
after the high-energy emission, even though the peak magnitude is 
much fainter. The brightest observed flare in $R$-band was 2 magnitudes 
in three minutes (see Figure~\ref{fig:KeckFlare}).  Much like the behavior
seen in X-rays (\S\ref{sec:XRT}), such dramatic optical variability at late
times is unlike anything seen before from an extragalactic GRB
optical afterglow.  Unfortunately none of our optical data directly
overlap the XRT light curve, making a direct correlation between the two
impossible.

Two months after the burst, the optical counterpart faded in R-band
to fainter than 26.0 and three months after the burst, faded in  
$i^{\prime}$-band to fainter than 24.5 (see Table~\ref{tab:kecklc}).

\section{A Near Infrared Counterpart}
\label{sec:NIR}

Given the large line-of-sight extinction, we undertook late-time NIR imaging
at a variety of facilities to search for a quiescent counterpart to 
\event.  The results of our campaign are summarized in Table \ref{tab:NIR}.

In detail, 
we observed the field of \event\ with the Near InfraRed Imager and 
spectrograph 
(NIRI; \citealt{hji+03}) mounted on the 10-m Gemini North telescope on two
occasions.  On 2007 June 19 we obtained $18 \times 60$\,s images in 
the \textit{K}-band under exquisite seeing ($\sim 0.4\arcsec$) and
photometric conditions.  The observations on UT 2007
 July 15 suffered from poor seeing and clouds.

On UT 2007 June 21, starting 13:10,  we 
observed the transient with Laser Guide Star Adaptive Optics (LGS-AO;
\citealt{wcj+06,vbl+06}) on the Keck II telescope and the Near-Infrared Camera
2 (NIRC2).  A total of 17 images were obtained, each consisting of three 20s 
co-added integrations, in the $K^{\prime}$ filter using the wide-angle camera.
We also obtained further late-time observations on UT 2007 Sep 21 and UT 2007 Sep 30. 

Finally, \textit{J-} and \textit{H-}band images were obtained with the 
Wide-Field Infrared Camera (WIRC; \citealt{weh+03}) mounted on the 
Palomar Hale 200-inch (P200) telescope.  
Thirty four images each with integration time of 30 s were taken in each filter
on the night of UT 21 June 2007.

All but the LGS data were processed with standard IRAF routines. Custom 
routines in {\it Python} and {\it IDL} (written by JSB and LP) were used 
for the LGS-AO reductions;  a custom distortion correction (obtained by PBC
\footnote{http://www2.keck.hawaii.edu/inst/n2TopLev/post\_observing/dewarp/}) 
was applied to the LGS-AO imaging.
We created an astrometric solution using our Gemini/NIRI $K$-band image
from the night of June 19 relative to about fifty point sources from the
2-$\mu$m All-Sky Survey(2MASS; \citealt{scs+06}).  
The resulting RMS positional uncertainty was 0.125\arcsec\ 
in right ascension and 0.098\arcsec\ in declination.  This NIRI $K$-band
image was then used to create a catalog of about one hundred point sources 
for astrometric matching with all other images.  The NIRI $K$-band 
image was chosen because of the excellent seeing conditions 
($\sim 0.4\arcsec$) and the larger field of view in comparison to
NIRC2.  Typical RMS positional uncertainties relative
to the reference image were $\approx 0.07\arcsec$ in each coordinate.
Using these astrometric solutions, we determine a position for the optical
transient in the Keck $R$-band flares of $\alpha= 19^{\mathrm{h}}
55^{\mathrm{m}} 09\farcs646, \delta= +26^{\circ}
14\arcmin 05\farcs62$ (J2000.0).

\begin{figure*}[htb]
\centerline{\psfig{file=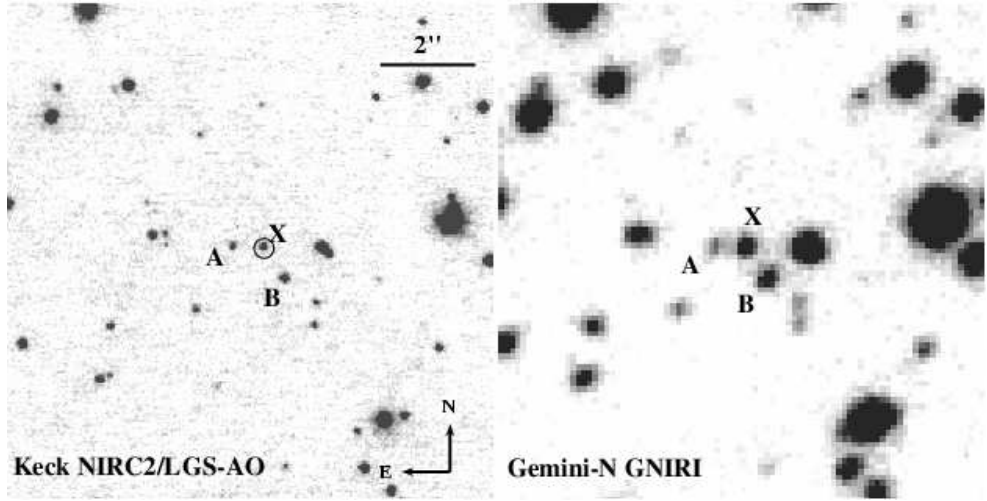,width=13cm}}
\caption[]{\small A $K$-band image of the field of \event\ obtained with
NIRC-2 imager behind the 
Keck~II Laser Guide Star (LGS) system (left; 2007 June 21) and NIRI imager
on the Gemini-North telescope (right; 2007 June 19).  The 2-$\sigma$ error
circle of the optical transient (taken from our \textit{LRIS} imaging)
is shown as a black circle overlaid on the LGS image.  Clearly we can
identify the object marked as 'X' as the NIR counterpart of \event.
}
\label{fig:NIRImages}
\end{figure*}

Despite the presence of two nearby objects ($A$ and
$B$), our astrometric accuracy is sufficient to unambiguously identify th $K$-band counterpart to \event\ ($X$ in Figure \ref{fig:NIRImages}).
Using the LGS-AO/NIRC2 image, we find that the location of this NIR counterpart 
is $\alpha= 19^{\mathrm{h}} 55^{\mathrm{m}} 9\farcs649$, $\delta = +26^{\circ}
14\arcmin 5\farcs65$ (J2000.0), with an uncertainty of 100\,mas in each
direction.  

Due to the crowded field, PSF-matched photometry
was performed on all images using the IRAF \texttt{DAOPHOT} package.
We summarize our NIR observations in Table \ref{tab:NIR}.  
%The $K$-band counterpart shows marginal evidence for fading 
%between the NIRI and LGS observations (Table \ref{tab:NIR}); our 
%photometric uncertainty may have been underestimated given some degree
%of contamination in the NIRI images.  
For reference, the $RIJHK_{\mathrm{s}}$
magnitudes of two extremely nearby objects $A$ and $B$ are 
provided in Table \ref{tab:ABphot}. Our late-time data, over three and 
a half months after the burst, constrains the quiescent counterpart to be 
fainter than $K^{\prime} >$ 21.5.

\begin{deluxetable}{llll}
  \tabletypesize{\footnotesize}
  \tablecaption{NIR Observations of \event}
  \tablecolumns{4}
  \tablewidth{0pc}
  \tablehead{\colhead{Epoch} & \colhead{Facility} & \colhead{Filter} &
             \colhead{Magnitude} \\
             \colhead{(2007 UT)} & & &
            }
  \startdata
    Jun 19.549 & Gemini-N/NIRI & $K$ & $19.30 \pm 0.23$ \\
    Jun 21.220 & Keck~II/LGS-AO+NIRC2 & $K^{\prime}$ & $19.83 \pm 0.15$\\
    Jul 15.309 & Gemini-N/NIRI & $K$ & $> 19.5$ \\
    Jun 21.352 & P200/WIRC & $J$ & $> 20.5$ \\
    Jun 21.400 & P200/WIRC & $H$ & $> 19.5$ \\
    Sep 21.632 & Keck~II/LGS-AO+NIRC2 & $K^{\prime}$ & $> 20.3$ \\
    Sep 30.264 & Keck~II/LGS-AO+NIRC2 & $K^{\prime}$ & $> 21.5$ \\    
  \enddata
  \tablecomments{Errors quoted are 1-$\sigma$ photometric and instrumental
    errors summed in quadrature.  Upper limits quoted are 3-$\sigma$.  No
    correction has been made for the large line-of-sight extinction.}
\label{tab:NIR}
\end{deluxetable}

\begin{deluxetable}{lllll}
  \tabletypesize{\footnotesize}
  \tablecaption{Photometry of Nearby Contaminating Sources $A$ and $B$}
  \tablecolumns{5}
  \tablewidth{0pc}
  \tablehead{\colhead{Epoch} & \colhead{Facility} & \colhead{Filter} & 
             \colhead{Magnitude} & \colhead{Magnitude} \\
             \colhead{(2007 UT)} & & & Source A & Source B
            }
  \startdata
    Jun 15.594 & Keck~I/LRIS & $R$ & $> 25.0$ & $> 25.0$ \\
    Aug 13.336 & Keck~I/LRIS & $R$ & $> 26.0$ & $> 26.0$ \\
    Jun 15.517 & Keck~I/LRIS & $I$ & $24.83 \pm 0.21$ & $24.93 \pm 0.22$ \\
    Jun 21.352 & P200/WIRC & $J$ & $> 20.5$ & $> 20.5$ \\
    Jun 21.400 & P200/WIRC & $H$ & $> 19.5$ & $> 19.5$ \\
    Jun 21.220 & Keck~II/LGS & $K^{\prime}$ & $20.30 \pm 0.16$ & $19.44 \pm 0.14$ \\

  \enddata
  \tablecomments{ Source B is $471\pm 22\,$mas West and
    $670\pm 22\,$mas South of \event\ . In images with poorer angular
    resolution, stars A and B may contaminate the photometry of the transient
    (i.e.~our NIRI imaging).}
\label{tab:ABphot}
\end{deluxetable}

\section{Search for a radio counterpart}
\label{sec:Radio}

On 2007 June 15 we undertook 
Very Large Array (VLA)\footnote{The National Radio Astronomy 
Observatory is a facility of the National Science Foundation operated under 
cooperative agreement by Associated Universities, Inc.} 
observations of \event.  The observations were obtained in $2\times 50\,$MHz
bands around 8.46\,GHz and lasted about an hour. 

We observed 1956+283 (a phase calibrator) for 0.8 minutes and then
switched to \event\ for 4.8 minutes. The sequence ended with a
6-minute observation of 0137+331 (3C48; flux calibrator).

Data were analyzed using the Astronomical Image Processing System (AIPS)
software of National Radio Astronomy Observatory (NRAO).  VLA
antennas N16, W64, E72 and W48 and baseline combinations EVLA
antennas E16 W24, N64, W40, E56, W48 and N40  were flagged.  In
total, flagging resulted in a loss of about 100 baselines.

Owing to the VLA being in the ``A'' configuration, we obtained
excellent image resolution of $0.42\arcsec\times 0.21\arcsec$. However,
\event\ was not detected and we get an upper limit of $7.3\pm 31.5\,\mu$Jy.

\section{Archival Observations}
\label{sec:Archival}

%RA=298.8046, Dec=26.2556
%l=63.3, b=-1.0 

A query of the {\it Simbad} database reveals no catalogued 
object within the BAT localization.  
The {\it INTEGRAL} observatory conducts regular scans of the Galactic plane 
and, in addition, performed several long pointed observations of the 
field around \event. Over the past four years, this field 
has been observed with the IBIS instrument to total 1.5 Ms being within 
its fully-coded field-of-view (FCFOV, $9^{\circ} \times 9^{\circ}$) and up 
to 2.5 Ms being within the partially coded field-of-view ($29^{\circ} 
\times 29^{\circ}$). 
The efficiency of observations within the FCFOV falls to zero at the 
field's edge. The coverage of these 4 years by observations was 
non-uniform with the maximum exposure reached in the fall of 2006 (for 
FCFOV).

There is no reported source close to the transient's position in the 
recent IBIS/ISGRI soft gamma-ray catalogs \citep{krl+07,bmb+07}.
We have also reanalyzed the archival data of {\it INTEGRAL} and 
failed to detect the source. A 4-$\sigma$ limit of 0.9 mCrab in the 
18--45 keV band (or 0.8 mCrab in the 17--60 keV band) has been received 
(flux of 1 mCrab corresponds to $1.1$ and $1.4\times10^{-11}$ erg 
cm$^{-2}$ s$^{-1}$ in these bands respectively for a source with Crab-like 
spectrum). There was also no source detected on a time scale of one 
individual pointing (2.0--3.6 ksec). We derive a 4-$\sigma$ limit 
of $\sim$20 mCrab.

Spitzer observed the position of \event\ during the Galactic Legacy
Infrared Mid-Plane Survey Extraordinaire (GLIMPSE) on 2004 Oct 31st.
Conservative upper limits for a source at the $K$-band position are
280,~350,~1700 and 6350 $\mu$Jy at 3.6,~4.5,~5.8 and 8.0$\mu$m respectively.

\section{Basic Considerations: Distance, Energetics and Radius}
\label{sec:Energetics}

The fluence, the hardness and the duration of GRB\,070610 are not atypical 
of GRBs.  However, \event\ is an atypical afterglow in the X-ray
band (\S\ref{sec:XRT}). The optical counterpart is also atypical.
Correcting for the total interstellar extinction along the line of sight, 
the apparent $i^{\prime}$-band magnitude of the optical transient is as 
bright as $\sim$13\,mag more than a day after the burst trigger -- there is 
no other optical extragalactic afterglow as bright at such late times.

The issue that faces us is quite simple: is GRB\,070610 related to
\event?
For extragalactic long-duration GRBs that the \Swift-XRT was able to observe
within an hour of the burst trigger, the overwhelming majority
have a detected X-ray afterglow.  

We therefore consider it unlikely that GRB\,070610 arises from a
background (i.e.~extragalactic) event. The spatial and temporal
coincidence of GRB\,070610 and \event\ suggest that these are strongly
related.  If so, the event is of Galactic origin. Accepting this association we 
turn our attention to the fundamental parameters characterizing
\event.

The extinction estimate based on the full X-ray spectrum 
excluding the flare (see Table~\ref{tab:XRTSpectrum})
corresponds to $E(B-V)$ = 1.0--1.5 mag (Based on optical spectral 
classification of nearby stars, we find $E(B-V) \sim$ 1.1).
Assuming $R=3.1$, this corresponds to $A_{K}$ = 0.4--0.5 mag.
The infrared $K^{\prime}$-band magnitude of the NIR counterpart 
\event\ is no brighter than $\sim 21.5$
(Table~\ref{tab:NIR}). From late-time optical observations, we also
know that R $>$ 26.0 (Table~\ref{tab:kecklc}). Since the farthest distance
for a star in the disk of our galaxy is 30 kpc, we get extinction-corrected 
absolute magnitude of M$_{K} > $3.6 and M$_{R} > $4.8. This clearly
rules out the luminosity class of giants and supergiants \citep{cox+00}. 
This also constrains the spectral type to be cooler than G8 \citep{kh+07}.    
If we assume a distance of 10 kpc, we can further constrain it to a 
spectral type cooler than M3. We also note here that in the case of another 
black-hole binary, XTE J1550-564, \citet{ogv+02} find that the gravity 
is lower than the nominal gravity for dwarfs.

The prompt $\gamma$-ray burst peak flux is $5
\times 10^{-8}\ {\rm erg\ cm^{-2} s^{-1}}$, the peak X-ray flare flux
is twice as faint and the mean flux over the first week is
approximately a factor of $10^{4}$ fainter than the burst peak flux.
These translate to the following isotropic luminosities: $6 \times
10^{38}\ d_{10}^2\ {\rm erg\ s^{-1}}$, $3 \times 10^{38}\ d_{10}^2\ {\rm
erg\ s^{-1}}$ and $5 \times 10^{34}\ d_{10}^2\ {\rm erg\ s^{-1}}$.
  
The prompt gamma-rays can constrain the radius of the emission. The
BAT burst duration of several seconds (see Figure 1) puts an upper
limit on the size of the emitting region (along the line of sight)
to be smaller than $\sim 10^{11} \beta_c$ cm, where $\beta_c$ is the
causal speed in units of the light speed (i.e, the speed in which
information, such as sound, travels). Since we expect $\beta_c \ll
1$ in non-compact objects (e.g., main sequence stars) the source of
the prompt gamma-rays is a black hole, a neutron star or a white
dwarf (the sound crossing time of the latter is seconds).
 
On the other hand, the non-thermal gamma-rays can be used to put a
lower limit on the emission radius. If the gamma-ray spectrum
continues to high energy ($E > m_ec^2$) then pair production opacity
starts playing a role. Using the formulation of \citet{ls01}
and assuming a non-relativistic source we find that the size
of the emitting region has to be $\gtrsim L\sigma_T /(0.1\pi m_ec^3)
\approx 10^9d_{10}^2$ cm, where the approximated numerical
factor(taken here as  $0.1\pi$) depends on the radiation spectrum
and the geometry of the source \citep{s87}. This radius implies that
if the engine of the burst is a neutron star or a black hole then
the observed gamma-rays are produced far from the engine by
(possibly relativistic) ejecta.

\section{A Curious Galactic Transient}

With a compact object (\S\ref{sec:Energetics}) and a fainter than K-dwarf
companion, \event\ is likely a binary system.
We now turn our attention to investigate the mechanism powering the unusual
emission, with an emphasis on identifying analogous systems in our Galaxy.

At first blush, soft $\gamma$-ray repeater (SGR) flares appear to be a viable
model for \event. SGR exhibits hard X-ray flares with durations 
ranging from 0.1\,s to 10\,s and isotropic luminosities of $10^{46}$\,erg
\citep{afg+01,hbs+05}.  Furthermore, variable X-ray afterglows have been 
detected after several SGR outbursts (see e.g.~\citealt{wt06}).

However, this interpretation has several problems.
First, SGR flares
lasting longer than 1\,s, dubbed ``intermediate'' SGR flares, have an energy
release $\approx 10^{41}$\,erg, two orders of magnitude larger than our upper
limit for \event\ \citep{wt06}.  Second, pulsations are typically observed in
SGR flare X-ray afterglows at the neutron star spin rate.  We see no evidence
for pulsations from \event\ (though constrained by 
the 2.5\,s sampling interval).  Finally, no
known SGR has a companion. 
If \event\ were caused by an SGR flare, a cooler than K-dwarf
companion would make \event\ the first binary magnetar.

Unlike SGR flares, the remaining possibilities are ultimately powered by
accretion instead of magnetic activity (\citealt{apb+03} provide a 
comprehensive review of such transients in the hard X-ray sky).
Cygnus X-1, a black hole binary with a supergiant companion, exhibits 
hard x-ray outbursts (\citealt{sbp+01}; \citealt{gaf+03}). 
The {\it INTEGRAL} mission has identified a class of
bright hard X-ray transients, the so-called Supergiant Fast X-ray
Transients (SFXT; \citealt{nst+07}). However, these events
are relatively soft, and have timescales of ~$10^3\,$s or longer. Furthermore,
the super-giant donor is an essential part of the SFXT story -- 
the X-ray flares arise from accretion of ``blobs'' in the wind of the
supergiant star. The faintness of the quiescent counterpart convincingly
rules out the giant and supergiant scenarios. 

The high peak luminosity strongly suggests an event like 
CI\,Cam (see \citealt{bdv+99}). However, this too is a questionable
analog for the reasons of the lack of a bright optical/NIR counterpart and 
also the short flare duration.  For the same reasons, the analogy to 
A0538$-$66 (the well known Be-pulsar X-ray binary in the LMC) can also be 
ruled out.

The bursting pulsar GRO\,J1744$-$28 shares some properties with those 
from GRB\,070610.  From 1995--1997, thousands of bursts
were detected by {\it BATSE} out to $> 60$\,keV \citep{kvf+96,wkv+99}.  The spectrum
of the bursts in {\it BATSE} and {\it RXTE} was adequately modeled by a thermal 
bremsstrahlung model having $kT \sim 10$\,keV; burst durations 
were approximately 10\,s.  
GRO\,J1744$-$28 consists of a neutron star in an 11.8\,d orbit with a
low mass companion \citep{fkn+96}.  However, unlike \event, there is no
evidence for a highly variable optical emission associated with these bursts.
Also the high-energy bursts from GRO\,J1744$-$28 are highly repetitive. Searches
for other episodes of emission from GRB\,070610 did not yield any obvious
candidates either in the BAT data or in the extensive {\it INTEGRAL} survey 
(\S\ref{sec:Archival}) and the Interplanetary Network (\S\ref{sec:Implications}).

The best analog to the X-ray and optical emission from \event\ is 
V4641~Sgr \citep{mpe+07} -- a transient which has been recognized by
several authors as being one of the fastest transients in the
hard X-ray \citep{ikb+00,ukw+02,apb+03}.
V4641~Sgr came to the attention of astronomers through a major
outburst in 1999 (see \citealt{ikb+00}). We now infer that this object
is a binary consisting of a B9 III star orbiting a  9\,$M_\odot$
black hole \citep{okv+01}.  The system exhibited strong and fast
X-ray and optical variability -- similar to what we see in \event.  

Rapid ($<$ 100\,s) and intense (modulation index,
$S=\langle f\rangle/\Delta(f)\gtrsim 10$; here $f$ is the X-ray flux and
$\Delta f$ is the variability in $f$) but with mean X-ray luminosity
$\langle L\rangle$ that is well below Eddington flux mark V4641~Sgr from the
other black hole binaries \citep{rgc+02}.  The classical black hole LMXBs such as
A0620$-$00 exhibit X-ray novae with peak super-Eddington flux and
a decline over a month (see reviews by \citealt{ts96,rm06}).
Micro-quasars such as GRS~1915+10 exhibit
intense variations (with $S$ approaching ten) but only when 
$\langle{L}\rangle$
is extremely high, $\langle{L}\rangle\sim 10^{39}\ $erg\ s$^{-1}$ (e.g.
\citealt{bmk+97,mmr99}).

The first difference between \event\ and V4641~Sgr is the donor star:
\event\ has a cool dwarf donor, while V4641~Sgr has a B9 giant 
donor. We suggest that
the distinctive variability of V4641~Sgr arises from
the black hole companion and has less to do with the nature
of the donor star. This conjecture would allow us to infer that
the compact object in \event\ is also a black hole.
While V4641~Sgr seems to be the closest
event we have to \event, it is clear that no perfect
analog to \event\ exists.   In particular, there has been
no report of a burst of gamma-rays from V4641~Sgr. However, the
absence could be due to the short duration duty cycle of
the gamma-ray bursts.

With two similar objects in hand---V4641~Sgr and GRB\,070610---we 
now have the luxury of defining a new
class of transients: fast X-ray novae which, in addition to the
rapid X-ray and optical variability but at sub-Eddington luminosities,
are also (apparently) marked by GRB-like bursts.

What differentiates fast X-ray novae from the regular X-ray novae?
Regular X-ray novae are essentially black hole binaries undergoing the
equivalent of dwarf novae i.e.~instabilities in the disk. 
During the major burst of 1999, V4641~Sgr exhibited radio emission and
relativistic motion \citep{hrh+00}. The radio flux of V4641~Sgr declined 
very steeply initially; from 360 mJy at 8.3 GHz to 30 mJy at 8.46 GHz in 
one day \citep{hrh+00}. According to \citealt{okv+01}, 
the distance to V4641~Sgr is 7.4--12.3 kpc and  
the apparent expansion velocity is $ > 9.5c$ (assuming the lowest
proper motion estimate from \citealt{hrh+00}) -- 
making V4641~Sgr the most relativistic of Galactic sources. This 
suggests that perhaps the key difference between fast X-ray novae and the
regular X-ray novae is the speed at which the ejecta is emitted.
Unfortunately, neither was GRB070610 as bright as V4641~Sgr in the optical
and X-ray immediately after the flare nor were our radio observations 
undertaken promptly after the detection of GRB 070610 to verify this 
hypothesis. 

\section{Implications: Galactic GRBs}
\label{sec:Implications}

Spurred by the connection between \event\ with a Galactic
transient we investigated whether this source or its analog V4641~Sgr
emitted bursts of gamma-rays in the past.  We have constructed a list of 
1211 GRBs detected by the IPN~\citep{hbk+99}, whose
position is constrained by at least one annulus with semi-width
smaller than $0.5$~deg.  This catalog contains events observed from
1990 November 12, to 2005 October 31 (see Ofek 2007 for more details).
We did not find any IPN GRB that coincides with either of these positions.

We also searched for \Swift-BAT sub-threshold events which are
consistent with the positions of V4641~Sgr and \event.  There
is no BAT sub-threshold event within 5\arcmin\ from the location of V4641~Sgr.  
But, we find an event at a signal-to-noise ratio (SNR)
of 5.0 located at RA=298.77$^\circ$, Dec=+26.221$^\circ$ (1.8\arcmin\ 
from the position of GRB\,070610) and occurring on UT 2006 Nov 17.7812.
However, adjusted for the approximate number of times this field has been
observed, the significance drops below 2-$\sigma$.  We consider it likely this
sub-threshold event is nothing more than a statistical fluctuation.

V4641~Sgr has been undergoing major bursts approximately every
two years (see \citealt{uki+04}). The absence of a detection of a
gamma-ray burst could simply
be due to lack of coverage or that not all such bursting activity
are preceded by a burst of gamma-rays. 

Another possible member of this class of fast X-ray novae is 
XTEJ1901+014~\citep{klg+07} which is potentially associated with 
GRB020406~\citep{Remillard02_ATEL88}. 

Nonetheless, it is reasonable to speculate that a burst similar
to \event\ occurs in our Galaxy, say, every decade. This alone
immediately makes \event\ and related events as the most common 
of long-duration gamma-ray bursts. (The mean time between cosmological GRBs in
our Galaxy is no smaller than $10^5\,$yr). 

Several hundred years ago, optical astronomers put all new apparitions
of stars as {\it novae stella.} Over the past century astronomers
have shown that {\it novae stella} split into three distinctly different
phenomena: 
novae, supernovae of type Ia and core collapse supernovae. The novae,
in turn, are divided into five families which arise from instabilities
in the accretion disk feeding a white dwarf, neutron star or a black hole
and on the surfaces of white dwarfs and neutron stars.

History is repeating itself. Only thirty years ago, astronomers referred to
all bursts of gamma-ray radiation as GRBs. 
Over the last decade
astronomers have established SHBs and LSBs to be of cosmological origin
\citep{mdk+97,gso+05,bpp+06,ffp+05}
and reasonably established their origin: 
coalescence of compact objects and deaths of massive stars
respectively. 

However, fissures are already developing.
Recently, hypergiant flares from magnetars in our
own Galaxy and nearby galaxies have been found to contaminate
the SHB sample. The Galactic rate of the hypergiant
flares is likely $10^{-3}\,$yr$^{-1}$ \citep{ofek07} much
larger than the estimated Galactic SHB rate of $10^{-6}\,$yr$^{-1}$
(\citealt{ngf06};~\citealt{gp+06}). 

Our galaxy has at least two fast X-ray novae systems (V4641~Sgr and
\event). The rate of GRB 070610-like events (with no assumption about
beaming) is likely to be about $3.5_{-2.9}^{+8.0}\,$yr$^{-1}$ which is 
five orders of magnitude larger than the estimated cosmological GRBs rate. 
However, these events are yet unobservable outside our galaxy with the 
current limitation in sensitivity of high energy detectors. As usual the 
meekest events dominate the demography. 

\begin{acknowledgements}
We thank M. van Kerkwijk for help with Keck observations and
discussions. We thank J. Cohen, J. Simon, A. Kraus, M. Muno, 
E. S. Phinney and R. Narayan for valuable discussions. We also acknowledge 
D. Law, T. Treu and P. Marshall. As always we are grateful to the selfless 
librarians and astronomers who maintain the {\it Simbad} database. 
M.~M.~K.~thanks the Gordon and Betty Moore Foundation for
support with the George Ellory Hale Fellowship.
S.~B.~C.~and A.~M.~S.~are supported by a NASA
Graduate Student Research Fellowship. JSB is a Sloan Research 
Fellow and is partially supported by a Hellman Faculty Award. 
P.~C.~S.~is supported by a Jansky Fellowship.  E.~B.~is supported 
by a Hubble fellowship.   
GRB research at Caltech is supported in part by grants from NSF (AST program)
and NASA (Swift and HST mission).

\end{acknowledgements}

\bibliographystyle{apj}
%\bibliography{ms}

\clearpage

\begin{thebibliography}{74}
\expandafter\ifx\csname natexlab\endcsname\relax\def\natexlab#1{#1}\fi

\bibitem[{{Aptekar} {et~al.}(2001){Aptekar}, {Frederiks}, {Golenetskii},
  {Il'inskii}, {Mazets}, {Pal'shin}, {Butterworth}, \& {Cline}}]{afg+01}
{Aptekar}, R.~L., {Frederiks}, D.~D., {Golenetskii}, S.~V., {Il'inskii}, V.~N.,
  {Mazets}, E.~P., {Pal'shin}, V.~D., {Butterworth}, P.~S., \& {Cline}, T.~L.
  2001, \apjs, 137, 227

\bibitem[{{Arefiev} {et~al.}(2003){Arefiev}, {Priedhorsky}, \&
  {Borozdin}}]{apb+03}
{Arefiev}, V.~A., {Priedhorsky}, W.~C., \& {Borozdin}, K.~N. 2003, \apj, 586,
  1238

\bibitem[{{Barthelmy} {et~al.}(2005){Barthelmy}, {Barbier}, {Cummings},
  {Fenimore}, {Gehrels}, {Hullinger}, {Krimm}, {Markwardt}, {Palmer},
  {Parsons}, {Sato}, {Suzuki}, {Takahashi}, {Tashiro}, \& {Tueller}}]{bbc+05}
{Barthelmy}, S.~D., {Barbier}, L.~M., {Cummings}, J.~R., {Fenimore}, E.~E.,
  {Gehrels}, N., {Hullinger}, D., {Krimm}, H.~A., {Markwardt}, C.~B., {Palmer},
  D.~M., {Parsons}, A., {Sato}, G., {Suzuki}, M., {Takahashi}, T., {Tashiro},
  M., \& {Tueller}, J. 2005, Space Science Reviews, 120, 143

\bibitem[{{Belloni} {et~al.}(1999){Belloni}, {Dieters}, {van den Ancker},
  {Fender}, {Fox}, {Harmon}, {van der Klis}, {Kommers}, {Lewin}, \& {van
  Paradijs}}]{bdv+99}
{Belloni}, T., {Dieters}, S., {van den Ancker}, M.~E., {Fender}, R.~P., {Fox},
  D.~W., {Harmon}, B.~A., {van der Klis}, M., {Kommers}, J.~M., {Lewin},
  W.~H.~G., \& {van Paradijs}, J. 1999, \apj, 527, 345

\bibitem[{{Belloni} {et~al.}(1997){Belloni}, {Mendez}, {King}, {van der Klis},
  \& {van Paradijs}}]{bmk+97}
{Belloni}, T., {Mendez}, M., {King}, A.~R., {van der Klis}, M., \& {van
  Paradijs}, J. 1997, \apjl, 488, L109+

\bibitem[{{Bird} {et~al.}(2007){Bird}, {Malizia}, {Bazzano}, {Barlow},
  {Bassani}, {Hill}, {B{\'e}langer}, {Capitanio}, {Clark}, {Dean}, {Fiocchi},
  {G{\"o}tz}, {Lebrun}, {Molina}, {Produit}, {Renaud}, {Sguera}, {Stephen},
  {Terrier}, {Ubertini}, {Walter}, {Winkler}, \& {Zurita}}]{bmb+07}
{Bird}, A.~J., {Malizia}, A., {Bazzano}, A., {Barlow}, E.~J., {Bassani}, L.,
  {Hill}, A.~B., {B{\'e}langer}, G., {Capitanio}, F., {Clark}, D.~J., {Dean},
  A.~J., {Fiocchi}, M., {G{\"o}tz}, D., {Lebrun}, F., {Molina}, M., {Produit},
  N., {Renaud}, M., {Sguera}, V., {Stephen}, J.~B., {Terrier}, R., {Ubertini},
  P., {Walter}, R., {Winkler}, C., \& {Zurita}, J. 2007, \apjs, 170, 175

\bibitem[{{Bloom} {et~al.}(2006){Bloom}, {Prochaska}, {Pooley}, {Blake},
  {Foley}, {Jha}, {Ramirez-Ruiz}, {Granot}, {Filippenko}, {Sigurdsson},
  {Barth}, {Chen}, {Cooper}, {Falco}, {Gal}, {Gerke}, {Gladders}, {Greene},
  {Hennanwi}, {Ho}, {Hurley}, {Koester}, {Li}, {Lubin}, {Newman}, {Perley},
  {Squires}, \& {Wood-Vasey}}]{bpp+06}
{Bloom}, J.~S., {Prochaska}, J.~X., {Pooley}, D., {Blake}, C.~H., {Foley},
  R.~J., {Jha}, S., {Ramirez-Ruiz}, E., {Granot}, J., {Filippenko}, A.~V.,
  {Sigurdsson}, S., {Barth}, A.~J., {Chen}, H.-W., {Cooper}, M.~C., {Falco},
  E.~E., {Gal}, R.~R., {Gerke}, B.~F., {Gladders}, M.~D., {Greene}, J.~E.,
  {Hennanwi}, J., {Ho}, L.~C., {Hurley}, K., {Koester}, B.~P., {Li}, W.,
  {Lubin}, L., {Newman}, J., {Perley}, D.~A., {Squires}, G.~K., \&
  {Wood-Vasey}, W.~M. 2006, \apj, 638, 354

\bibitem[{{Buccheri} {et~al.}(1983){Buccheri}, {Bennett}, {Bignami}, {Bloemen},
  {Boriakoff}, {Caraveo}, {Hermsen}, {Kanbach}, {Manchester}, {Masnou},
  {Mayer-Hasselwander}, {Ozel}, {Paul}, {Sacco}, {Scarsi}, \&
  {Strong}}]{bbb+83}
{Buccheri}, R., {Bennett}, K., {Bignami}, G.~F., {Bloemen}, J.~B.~G.~M.,
  {Boriakoff}, V., {Caraveo}, P.~A., {Hermsen}, W., {Kanbach}, G.,
  {Manchester}, R.~N., {Masnou}, J.~L., {Mayer-Hasselwander}, H.~A., {Ozel},
  M.~E., {Paul}, J.~A., {Sacco}, B., {Scarsi}, L., \& {Strong}, A.~W. 1983,
  \aap, 128, 245

\bibitem[{{Burrows} {et~al.}(2005){Burrows}, {Hill}, {Nousek}, {Kennea},
  {Wells}, {Osborne}, {Abbey}, {Beardmore}, {Mukerjee}, {Short}, {Chincarini},
  {Campana}, {Citterio}, {Moretti}, {Pagani}, {Tagliaferri}, {Giommi},
  {Capalbi}, {Tamburelli}, {Angelini}, {Cusumano}, {Br{\"a}uninger}, {Burkert},
  \& {Hartner}}]{bhn+05}
{Burrows}, D.~N., {Hill}, J.~E., {Nousek}, J.~A., {Kennea}, J.~A., {Wells}, A.,
  {Osborne}, J.~P., {Abbey}, A.~F., {Beardmore}, A., {Mukerjee}, K., {Short},
  A.~D.~T., {Chincarini}, G., {Campana}, S., {Citterio}, O., {Moretti}, A.,
  {Pagani}, C., {Tagliaferri}, G., {Giommi}, P., {Capalbi}, M., {Tamburelli},
  F., {Angelini}, L., {Cusumano}, G., {Br{\"a}uninger}, H.~W., {Burkert}, W.,
  \& {Hartner}, G.~D. 2005, Space Science Reviews, 120, 165

\bibitem[{{Cardelli} {et~al.}(1989){Cardelli}, {Clayton}, \& {Mathis}}]{ccm+89}
{Cardelli}, J.~A., {Clayton}, G.~C., \& {Mathis}, J.~S. 1989, \apj, 345, 245

\bibitem[{{Cenko} {et~al.}(2006){Cenko}, {Fox}, {Moon}, {Harrison}, {Kulkarni},
  {Henning}, {Guzman}, {Bonati}, {Smith}, {Thicksten}, {Doyle}, {Petrie},
  {Gal-Yam}, {Soderberg}, {Anagnostou}, \& {Laity}}]{cfm+06}
{Cenko}, S.~B., {Fox}, D.~B., {Moon}, D.-S., {Harrison}, F.~A., {Kulkarni},
  S.~R., {Henning}, J.~R., {Guzman}, C.~D., {Bonati}, M., {Smith}, R.~M.,
  {Thicksten}, R.~P., {Doyle}, M.~W., {Petrie}, H.~L., {Gal-Yam}, A.,
  {Soderberg}, A.~M., {Anagnostou}, N.~L., \& {Laity}, A.~C. 2006, \pasp, 118,
  1396

\bibitem[{{Chincarini} {et~al.}(2007){Chincarini}, {Moretti}, {Romano},
  {Falcone}, {Morris}, {Racusin}, {Campana}, {Guidorzi}, {Tagliaferri},
  {Burrows}, {Pagani}, {Stroh}, {Grupe}, {Capalbi}, {Cusumano}, {Gehrels},
  {Giommi}, {La Parola}, {Mangano}, {Mineo}, {Nousek}, {O'Brien}, {Page},
  {Perri}, {Troja}, {Willingale}, \& {Zhang}}]{cmr+07}
{Chincarini}, G., {Moretti}, A., {Romano}, P., {Falcone}, A.~D., {Morris}, D.,
  {Racusin}, J., {Campana}, S., {Guidorzi}, C., {Tagliaferri}, G., {Burrows},
  D.~N., {Pagani}, C., {Stroh}, M., {Grupe}, D., {Capalbi}, M., {Cusumano}, G.,
  {Gehrels}, N., {Giommi}, P., {La Parola}, V., {Mangano}, V., {Mineo}, T.,
  {Nousek}, J.~A., {O'Brien}, P.~T., {Page}, K.~L., {Perri}, M., {Troja}, E.,
  {Willingale}, R., \& {Zhang}, B. 2007, ArXiv Astrophysics e-prints

\bibitem[{{Cox}(2000)}]{cox+00}
{Cox}, A.~N. 2000, {Allen's astrophysical quantities} (Allen's astrophysical
  quantities, 4th ed.~Publisher: New York: AIP Press; Springer, 2000.~Editedy
  by Arthur N.~Cox.~ ISBN: 0387987460)

\bibitem[{{Dickey} \& {Lockman}(1990)}]{dl90}
{Dickey}, J.~M. \& {Lockman}, F.~J. 1990, \araa, 28, 215

\bibitem[{{Evans} {et~al.}(2007){Evans}, {Beardmore}, {Page}, {Tyler},
  {Osborne}, {Goad}, {O'Brien}, {Vetere}, {Racusin}, {Morris}, {Burrows},
  {Capalbi}, {Perri}, {Gehrels}, \& {Romano}}]{ebp+07}
{Evans}, P.~A., {Beardmore}, A.~P., {Page}, K.~L., {Tyler}, L.~G., {Osborne},
  J.~P., {Goad}, M.~R., {O'Brien}, P.~T., {Vetere}, L., {Racusin}, J.,
  {Morris}, D., {Burrows}, D.~N., {Capalbi}, M., {Perri}, M., {Gehrels}, N., \&
  {Romano}, P. 2007, \aap, 469, 379

\bibitem[{{Falcone} {et~al.}(2006){Falcone}, {Burrows}, {Lazzati}, {Campana},
  {Kobayashi}, {Zhang}, {M{\'e}sz{\'a}ros}, {Page}, {Kennea}, {Romano},
  {Pagani}, {Angelini}, {Beardmore}, {Capalbi}, {Chincarini}, {Cusumano},
  {Giommi}, {Goad}, {Godet}, {Grupe}, {Hill}, {La Parola}, {Mangano},
  {Moretti}, {Nousek}, {O'Brien}, {Osborne}, {Perri}, {Tagliaferri}, {Wells},
  \& {Gehrels}}]{fbl+06}
{Falcone}, A.~D., {Burrows}, D.~N., {Lazzati}, D., {Campana}, S., {Kobayashi},
  S., {Zhang}, B., {M{\'e}sz{\'a}ros}, P., {Page}, K.~L., {Kennea}, J.~A.,
  {Romano}, P., {Pagani}, C., {Angelini}, L., {Beardmore}, A.~P., {Capalbi},
  M., {Chincarini}, G., {Cusumano}, G., {Giommi}, P., {Goad}, M.~R., {Godet},
  O., {Grupe}, D., {Hill}, J.~E., {La Parola}, V., {Mangano}, V., {Moretti},
  A., {Nousek}, J.~A., {O'Brien}, P.~T., {Osborne}, J.~P., {Perri}, M.,
  {Tagliaferri}, G., {Wells}, A.~A., \& {Gehrels}, N. 2006, \apj, 641, 1010

\bibitem[{{Finger} {et~al.}(1996){Finger}, {Koh}, {Nelson}, {Prince},
  {Vaughan}, \& {Wilson}}]{fkn+96}
{Finger}, M.~H., {Koh}, D.~T., {Nelson}, R.~W., {Prince}, T.~A., {Vaughan},
  B.~A., \& {Wilson}, R.~B. 1996, \nat, 381, 291

\bibitem[{{Fox} {et~al.}(2005){Fox}, {Frail}, {Price}, {Kulkarni}, {Berger},
  {Piran}, {Soderberg}, {Cenko}, {Cameron}, {Gal-Yam}, {Kasliwal}, {Moon},
  {Harrison}, {Nakar}, {Schmidt}, {Penprase}, {Chevalier}, {Kumar}, {Roth},
  {Watson}, {Lee}, {Shectman}, {Phillips}, {Roth}, {McCarthy}, {Rauch},
  {Cowie}, {Peterson}, {Rich}, {Kawai}, {Aoki}, {Kosugi}, {Totani}, {Park},
  {MacFadyen}, \& {Hurley}}]{ffp+05}
{Fox}, D.~B., {Frail}, D.~A., {Price}, P.~A., {Kulkarni}, S.~R., {Berger}, E.,
  {Piran}, T., {Soderberg}, A.~M., {Cenko}, S.~B., {Cameron}, P.~B., {Gal-Yam},
  A., {Kasliwal}, M.~M., {Moon}, D.-S., {Harrison}, F.~A., {Nakar}, E.,
  {Schmidt}, B.~P., {Penprase}, B., {Chevalier}, R.~A., {Kumar}, P., {Roth},
  K., {Watson}, D., {Lee}, B.~L., {Shectman}, S., {Phillips}, M.~M., {Roth},
  M., {McCarthy}, P.~J., {Rauch}, M., {Cowie}, L., {Peterson}, B.~A., {Rich},
  J., {Kawai}, N., {Aoki}, K., {Kosugi}, G., {Totani}, T., {Park}, H.-S.,
  {MacFadyen}, A., \& {Hurley}, K.~C. 2005, \nat, 437, 845

\bibitem[{{French} {et~al.}(2007){French}, {Melady}, {Kubanek}, \&
  {Jelinek}}]{French07_GCN6500}
{French}, J., {Melady}, G., {Kubanek}, P., \& {Jelinek}, M. 2007, {GCN
  Circular} 6500

\bibitem[{{Gehrels} {et~al.}(2004){Gehrels}, {Chincarini}, {Giommi}, {Mason},
  {Nousek}, {Wells}, {White}, {Barthelmy}, {Burrows}, {Cominsky}, {Hurley},
  {Marshall}, {M{\'e}sz{\'a}ros}, {Roming}, {Angelini}, {Barbier}, {Belloni},
  {Campana}, {Caraveo}, {Chester}, {Citterio}, {Cline}, {Cropper}, {Cummings},
  {Dean}, {Feigelson}, {Fenimore}, {Frail}, {Fruchter}, {Garmire}, {Gendreau},
  {Ghisellini}, {Greiner}, {Hill}, {Hunsberger}, {Krimm}, {Kulkarni}, {Kumar},
  {Lebrun}, {Lloyd-Ronning}, {Markwardt}, {Mattson}, {Mushotzky}, {Norris},
  {Osborne}, {Paczynski}, {Palmer}, {Park}, {Parsons}, {Paul}, {Rees},
  {Reynolds}, {Rhoads}, {Sasseen}, {Schaefer}, {Short}, {Smale}, {Smith},
  {Stella}, {Tagliaferri}, {Takahashi}, {Tashiro}, {Townsley}, {Tueller},
  {Turner}, {Vietri}, {Voges}, {Ward}, {Willingale}, {Zerbi}, \&
  {Zhang}}]{gcg+04}
{Gehrels}, N., {Chincarini}, G., {Giommi}, P., {Mason}, K.~O., {Nousek}, J.~A.,
  {Wells}, A.~A., {White}, N.~E., {Barthelmy}, S.~D., {Burrows}, D.~N.,
  {Cominsky}, L.~R., {Hurley}, K.~C., {Marshall}, F.~E., {M{\'e}sz{\'a}ros},
  P., {Roming}, P.~W.~A., {Angelini}, L., {Barbier}, L.~M., {Belloni}, T.,
  {Campana}, S., {Caraveo}, P.~A., {Chester}, M.~M., {Citterio}, O., {Cline},
  T.~L., {Cropper}, M.~S., {Cummings}, J.~R., {Dean}, A.~J., {Feigelson},
  E.~D., {Fenimore}, E.~E., {Frail}, D.~A., {Fruchter}, A.~S., {Garmire},
  G.~P., {Gendreau}, K., {Ghisellini}, G., {Greiner}, J., {Hill}, J.~E.,
  {Hunsberger}, S.~D., {Krimm}, H.~A., {Kulkarni}, S.~R., {Kumar}, P.,
  {Lebrun}, F., {Lloyd-Ronning}, N.~M., {Markwardt}, C.~B., {Mattson}, B.~J.,
  {Mushotzky}, R.~F., {Norris}, J.~P., {Osborne}, J., {Paczynski}, B.,
  {Palmer}, D.~M., {Park}, H.-S., {Parsons}, A.~M., {Paul}, J., {Rees}, M.~J.,
  {Reynolds}, C.~S., {Rhoads}, J.~E., {Sasseen}, T.~P., {Schaefer}, B.~E.,
  {Short}, A.~T., {Smale}, A.~P., {Smith}, I.~A., {Stella}, L., {Tagliaferri},
  G., {Takahashi}, T., {Tashiro}, M., {Townsley}, L.~K., {Tueller}, J.,
  {Turner}, M.~J.~L., {Vietri}, M., {Voges}, W., {Ward}, M.~J., {Willingale},
  R., {Zerbi}, F.~M., \& {Zhang}, W.~W. 2004, \apj, 611, 1005

\bibitem[{{Gehrels} {et~al.}(2005){Gehrels}, {Sarazin}, {O'Brien}, {Zhang},
  {Barbier}, {Barthelmy}, {Blustin}, {Burrows}, {Cannizzo}, {Cummings}, {Goad},
  {Holland}, {Hurkett}, {Kennea}, {Levan}, {Markwardt}, {Mason}, {Meszaros},
  {Page}, {Palmer}, {Rol}, {Sakamoto}, {Willingale}, {Angelini}, {Beardmore},
  {Boyd}, {Breeveld}, {Campana}, {Chester}, {Chincarini}, {Cominsky},
  {Cusumano}, {de Pasquale}, {Fenimore}, {Giommi}, {Gronwall}, {Grupe}, {Hill},
  {Hinshaw}, {Hjorth}, {Hullinger}, {Hurley}, {Klose}, {Kobayashi},
  {Kouveliotou}, {Krimm}, {Mangano}, {Marshall}, {McGowan}, {Moretti},
  {Mushotzky}, {Nakazawa}, {Norris}, {Nousek}, {Osborne}, {Page}, {Parsons},
  {Patel}, {Perri}, {Poole}, {Romano}, {Roming}, {Rosen}, {Sato}, {Schady},
  {Smale}, {Sollerman}, {Starling}, {Still}, {Suzuki}, {Tagliaferri},
  {Takahashi}, {Tashiro}, {Tueller}, {Wells}, {White}, \& {Wijers}}]{gso+05}
{Gehrels}, N., {Sarazin}, C.~L., {O'Brien}, P.~T., {Zhang}, B., {Barbier}, L.,
  {Barthelmy}, S.~D., {Blustin}, A., {Burrows}, D.~N., {Cannizzo}, J.,
  {Cummings}, J.~R., {Goad}, M., {Holland}, S.~T., {Hurkett}, C.~P., {Kennea},
  J.~A., {Levan}, A., {Markwardt}, C.~B., {Mason}, K.~O., {Meszaros}, P.,
  {Page}, M., {Palmer}, D.~M., {Rol}, E., {Sakamoto}, T., {Willingale}, R.,
  {Angelini}, L., {Beardmore}, A., {Boyd}, P.~T., {Breeveld}, A., {Campana},
  S., {Chester}, M.~M., {Chincarini}, G., {Cominsky}, L.~R., {Cusumano}, G.,
  {de Pasquale}, M., {Fenimore}, E.~E., {Giommi}, P., {Gronwall}, C., {Grupe},
  D., {Hill}, J.~E., {Hinshaw}, D., {Hjorth}, J., {Hullinger}, D., {Hurley},
  K.~C., {Klose}, S., {Kobayashi}, S., {Kouveliotou}, C., {Krimm}, H.~A.,
  {Mangano}, V., {Marshall}, F.~E., {McGowan}, K., {Moretti}, A., {Mushotzky},
  R.~F., {Nakazawa}, K., {Norris}, J.~P., {Nousek}, J.~A., {Osborne}, J.~P.,
  {Page}, K., {Parsons}, A.~M., {Patel}, S., {Perri}, M., {Poole}, T.,
  {Romano}, P., {Roming}, P.~W.~A., {Rosen}, S., {Sato}, G., {Schady}, P.,
  {Smale}, A.~P., {Sollerman}, J., {Starling}, R., {Still}, M., {Suzuki}, M.,
  {Tagliaferri}, G., {Takahashi}, T., {Tashiro}, M., {Tueller}, J., {Wells},
  A.~A., {White}, N.~E., \& {Wijers}, R.~A.~M.~J. 2005, \nat, 437, 851

\bibitem[{{Golenetskii} {et~al.}(2003){Golenetskii}, {Aptekar}, {Frederiks},
  {Mazets}, {Palshin}, {Hurley}, {Cline}, \& {Stern}}]{gaf+03}
{Golenetskii}, S., {Aptekar}, R., {Frederiks}, D., {Mazets}, E., {Palshin}, V.,
  {Hurley}, K., {Cline}, T., \& {Stern}, B. 2003, \apj, 596, 1113

\bibitem[{{Guetta} \& {Piran}(2006)}]{gp+06}
{Guetta}, D. \& {Piran}, T. 2006, \aap, 453, 823

\bibitem[{{Hjellming} {et~al.}(2000){Hjellming}, {Rupen}, {Hunstead},
  {Campbell-Wilson}, {Mioduszewski}, {Gaensler}, {Smith}, {Sault}, {Fender},
  {Spencer}, {de la Force}, {Richards}, {Garrington}, {Trushkin}, {Ghigo},
  {Waltman}, \& {McCollough}}]{hrh+00}
{Hjellming}, R.~M., {Rupen}, M.~P., {Hunstead}, R.~W., {Campbell-Wilson}, D.,
  {Mioduszewski}, A.~J., {Gaensler}, B.~M., {Smith}, D.~A., {Sault}, R.~J.,
  {Fender}, R.~P., {Spencer}, R.~E., {de la Force}, C.~J., {Richards},
  A.~M.~S., {Garrington}, S.~T., {Trushkin}, S.~A., {Ghigo}, F.~D., {Waltman},
  E.~B., \& {McCollough}, M. 2000, \apj, 544, 977

\bibitem[{{Hodapp} {et~al.}(2003){Hodapp}, {Jensen}, {Irwin}, {Yamada},
  {Chung}, {Fletcher}, {Robertson}, {Hora}, {Simons}, {Mays}, {Nolan}, {Bec},
  {Merrill}, \& {Fowler}}]{hji+03}
{Hodapp}, K.~W., {Jensen}, J.~B., {Irwin}, E.~M., {Yamada}, H., {Chung}, R.,
  {Fletcher}, K., {Robertson}, L., {Hora}, J.~L., {Simons}, D.~A., {Mays}, W.,
  {Nolan}, R., {Bec}, M., {Merrill}, M., \& {Fowler}, A.~M. 2003, \pasp, 115,
  1388

\bibitem[{{Hurley} {et~al.}(2005){Hurley}, {Boggs}, {Smith}, {Duncan}, {Lin},
  {Zoglauer}, {Krucker}, {Hurford}, {Hudson}, {Wigger}, {Hajdas}, {Thompson},
  {Mitrofanov}, {Sanin}, {Boynton}, {Fellows}, {von Kienlin}, {Lichti}, {Rau},
  \& {Cline}}]{hbs+05}
{Hurley}, K., {Boggs}, S.~E., {Smith}, D.~M., {Duncan}, R.~C., {Lin}, R.,
  {Zoglauer}, A., {Krucker}, S., {Hurford}, G., {Hudson}, H., {Wigger}, C.,
  {Hajdas}, W., {Thompson}, C., {Mitrofanov}, I., {Sanin}, A., {Boynton}, W.,
  {Fellows}, C., {von Kienlin}, A., {Lichti}, G., {Rau}, A., \& {Cline}, T.
  2005, \nat, 434, 1098

\bibitem[{{Hurley} {et~al.}(1999){Hurley}, {Briggs}, {Kippen}, {Kouveliotou},
  {Meegan}, {Fishman}, {Cline}, \& {Boer}}]{hbk+99}
{Hurley}, K., {Briggs}, M.~S., {Kippen}, R.~M., {Kouveliotou}, C., {Meegan},
  C., {Fishman}, G., {Cline}, T., \& {Boer}, M. 1999, \apjs, 120, 399

\bibitem[{{in't Zand} {et~al.}(2000){in't Zand}, {Kuulkers}, {Bazzano},
  {Cornelisse}, {Cocchi}, {Heise}, {Muller}, {Natalucci}, {Smith}, \&
  {Ubertini}}]{ikb+00}
{in't Zand}, J.~J.~M., {Kuulkers}, E., {Bazzano}, A., {Cornelisse}, R.,
  {Cocchi}, M., {Heise}, J., {Muller}, J.~M., {Natalucci}, L., {Smith},
  M.~J.~S., \& {Ubertini}, P. 2000, \aap, 357, 520

\bibitem[{{Kalberla} {et~al.}(2005){Kalberla}, {Burton}, {Hartmann}, {Arnal},
  {Bajaja}, {Morras}, \& {P{\"o}ppel}}]{kbh+05}
{Kalberla}, P.~M.~W., {Burton}, W.~B., {Hartmann}, D., {Arnal}, E.~M.,
  {Bajaja}, E., {Morras}, R., \& {P{\"o}ppel}, W.~G.~L. 2005, \aap, 440, 775

\bibitem[{{Kann} {et~al.}(2007){Kann}, {Wilson}, {Schulze}, {Klose}, {Henze},
  {Ludwig}, {Laux}, \& {Greiner}}]{Kann07_GCN6505}
{Kann}, D.~A., {Wilson}, A.~C., {Schulze}, S., {Klose}, S., {Henze}, M.,
  {Ludwig}, F., {Laux}, U., \& {Greiner}, J. 2007, {GCN Circular} 6505

\bibitem[{{Karasev} {et~al.}(2007){Karasev}, {Lutovinov}, \&
  {Grebenev}}]{klg+07}
{Karasev}, D.~I., {Lutovinov}, A.~A., \& {Grebenev}, S.~A. 2007, Astronomy
  Letters, 33, 159

\bibitem[{{Klotz} {et~al.}(2007){Klotz}, {Boer}, {Atteia}, \&
  {Gendre}}]{Klotz07_GCN6513}
{Klotz}, K.~A., {Boer}, B.~M., {Atteia}, A.~J.~L., \& {Gendre}, G.~B. 2007,
  {GCN Circular} 6513

\bibitem[{{Kouveliotou} {et~al.}(1996){Kouveliotou}, {van Paradijs}, {Fishman},
  {Briggs}, {Kommers}, {Harmon}, {Meegan}, \& {Lewin}}]{kvf+96}
{Kouveliotou}, C., {van Paradijs}, J., {Fishman}, G.~J., {Briggs}, M.~S.,
  {Kommers}, J., {Harmon}, B.~A., {Meegan}, C.~A., \& {Lewin}, W.~H.~G. 1996,
  \nat, 379, 799

\bibitem[{{Kraus} \& {Hillenbrand}(2007)}]{kh+07}
{Kraus}, A.~L. \& {Hillenbrand}, L.~A. 2007, ArXiv e-prints, 708

\bibitem[{{Krivonos} {et~al.}(2007){Krivonos}, {Revnivtsev}, {Lutovinov},
  {Sazonov}, {Churazov}, \& {Sunyaev}}]{krl+07}
{Krivonos}, R., {Revnivtsev}, M., {Lutovinov}, A., {Sazonov}, S., {Churazov},
  E., \& {Sunyaev}, R. 2007, ArXiv Astrophysics e-prints

\bibitem[{{Lithwick} \& {Sari}(2001)}]{ls01}
{Lithwick}, Y. \& {Sari}, R. 2001, \apj, 555, 540

\bibitem[{{Markwardt} {et~al.}(2007){Markwardt}, {Pagani}, {Evans}, {Gavriil},
  {Kennea}, {Krimm}, {Landsman}, \& {Marshall}}]{mpe+07}
{Markwardt}, C.~B., {Pagani}, C., {Evans}, P., {Gavriil}, F.~P., {Kennea},
  J.~A., {Krimm}, H.~A., {Landsman}, W., \& {Marshall}, F.~E. 2007, The
  Astronomer's Telegram, 1102, 1

\bibitem[{{Metzger} {et~al.}(1997){Metzger}, {Djorgovski}, {Kulkarni},
  {Steidel}, {Adelberger}, {Frail}, {Costa}, \& {Frontera}}]{mdk+97}
{Metzger}, M.~R., {Djorgovski}, S.~G., {Kulkarni}, S.~R., {Steidel}, C.~C.,
  {Adelberger}, K.~L., {Frail}, D.~A., {Costa}, E., \& {Frontera}, F. 1997,
  \nat, 387, 878

\bibitem[{{Munari} {et~al.}(2005){Munari}, {Sordo}, {Castelli}, \&
  {Zwitter}}]{msc+05}
{Munari}, U., {Sordo}, R., {Castelli}, F., \& {Zwitter}, T. 2005, \aap, 442,
  1127

\bibitem[{{Muno} {et~al.}(1999){Muno}, {Morgan}, \& {Remillard}}]{mmr99}
{Muno}, M.~P., {Morgan}, E.~H., \& {Remillard}, R.~A. 1999, \apj, 527, 321

\bibitem[{{Nakar} {et~al.}(2006){Nakar}, {Gal-Yam}, \& {Fox}}]{ngf06}
{Nakar}, E., {Gal-Yam}, A., \& {Fox}, D.~B. 2006, \apj, 650, 281

\bibitem[{{Negueruela} {et~al.}(2007){Negueruela}, {Smith}, {Torrejon}, \&
  {Reig}}]{nst+07}
{Negueruela}, I., {Smith}, D.~M., {Torrejon}, J.~M., \& {Reig}, P. 2007, ArXiv
  e-prints, 704

\bibitem[{{Norris} {et~al.}(1996){Norris}, {Nemiroff}, {Bonnell}, {Scargle},
  {Kouveliotou}, {Paciesas}, {Meegan}, \& {Fishman}}]{nnb+96}
{Norris}, J.~P., {Nemiroff}, R.~J., {Bonnell}, J.~T., {Scargle}, J.~D.,
  {Kouveliotou}, C., {Paciesas}, W.~S., {Meegan}, C.~A., \& {Fishman}, G.~J.
  1996, \apj, 459, 393

\bibitem[{{Nousek} {et~al.}(2006){Nousek}, {Kouveliotou}, {Grupe}, {Page},
  {Granot}, {Ramirez-Ruiz}, {Patel}, {Burrows}, {Mangano}, {Barthelmy},
  {Beardmore}, {Campana}, {Capalbi}, {Chincarini}, {Cusumano}, {Falcone},
  {Gehrels}, {Giommi}, {Goad}, {Godet}, {Hurkett}, {Kennea}, {Moretti},
  {O'Brien}, {Osborne}, {Romano}, {Tagliaferri}, \& {Wells}}]{nkg+06}
{Nousek}, J.~A., {Kouveliotou}, C., {Grupe}, D., {Page}, K.~L., {Granot}, J.,
  {Ramirez-Ruiz}, E., {Patel}, S.~K., {Burrows}, D.~N., {Mangano}, V.,
  {Barthelmy}, S., {Beardmore}, A.~P., {Campana}, S., {Capalbi}, M.,
  {Chincarini}, G., {Cusumano}, G., {Falcone}, A.~D., {Gehrels}, N., {Giommi},
  P., {Goad}, M.~R., {Godet}, O., {Hurkett}, C.~P., {Kennea}, J.~A., {Moretti},
  A., {O'Brien}, P.~T., {Osborne}, J.~P., {Romano}, P., {Tagliaferri}, G., \&
  {Wells}, A.~A. 2006, \apj, 642, 389

\bibitem[{{Ofek}(2007)}]{ofek07}
{Ofek}, E.~O. 2007, \apj, 659, 339

\bibitem[{{Oke} {et~al.}(1995){Oke}, {Cohen}, {Carr}, {Cromer}, {Dingizian},
  {Harris}, {Labrecque}, {Lucinio}, {Schaal}, {Epps}, \& {Miller}}]{occ+95}
{Oke}, J.~B., {Cohen}, J.~G., {Carr}, M., {Cromer}, J., {Dingizian}, A.,
  {Harris}, F.~H., {Labrecque}, S., {Lucinio}, R., {Schaal}, W., {Epps}, H., \&
  {Miller}, J. 1995, \pasp, 107, 375

\bibitem[{{Orosz} {et~al.}(2002){Orosz}, {Groot}, {van der Klis}, {McClintock},
  {Garcia}, {Zhao}, {Jain}, {Bailyn}, \& {Remillard}}]{ogv+02}
{Orosz}, J.~A., {Groot}, P.~J., {van der Klis}, M., {McClintock}, J.~E.,
  {Garcia}, M.~R., {Zhao}, P., {Jain}, R.~K., {Bailyn}, C.~D., \& {Remillard},
  R.~A. 2002, \apj, 568, 845

\bibitem[{{Orosz} {et~al.}(2001){Orosz}, {Kuulkers}, {van der Klis},
  {McClintock}, {Garcia}, {Callanan}, {Bailyn}, {Jain}, \&
  {Remillard}}]{okv+01}
{Orosz}, J.~A., {Kuulkers}, E., {van der Klis}, M., {McClintock}, J.~E.,
  {Garcia}, M.~R., {Callanan}, P.~J., {Bailyn}, C.~D., {Jain}, R.~K., \&
  {Remillard}, R.~A. 2001, \apj, 555, 489

\bibitem[{{Pagani} {et~al.}(2007{\natexlab{a}}){Pagani}, {Barthelmy},
  {Cummings}, {Gehrels}, {Grupe}, {Holland}, {Kennea}, {Markwardt}, {Marshall},
  {O'Brien}, {Palmer}, {Parsons}, {Stamatikos}, \& {Vetere}}]{Pagani07_GCN6489}
{Pagani}, C., {Barthelmy}, S.~D., {Cummings}, J.~R., {Gehrels}, N., {Grupe},
  D., {Holland}, S.~T., {Kennea}, J.~A., {Markwardt}, C.~B., {Marshall}, F.~E.,
  {O'Brien}, P.~T., {Palmer}, D.~M., {Parsons}, A.~M., {Stamatikos}, M., \&
  {Vetere}, L. 2007{\natexlab{a}}, {GCN Circular} 6489

\bibitem[{{Pagani} {et~al.}(2007{\natexlab{b}}){Pagani}, {Racusin}, \&
  {Kennea}}]{Pagani07_GCN6506}
{Pagani}, C., {Racusin}, J.~L., \& {Kennea}, J.~A. 2007{\natexlab{b}}, {GCN
  Circular} 6506

\bibitem[{{Postigo} {et~al.}(2007){Postigo}, {Castro-Tirado}, \&
  {Aceituno}}]{Postigo07_GCN6501}
{Postigo}, A.~d.~U., {Castro-Tirado}, A.~J., \& {Aceituno}, F. 2007, {GCN
  Circular} 6501

\bibitem[{{Remillard} \& {McClintock}(2006)}]{rm06}
{Remillard}, R.~A. \& {McClintock}, J.~E. 2006, \araa, 44, 49

\bibitem[{{Remillard} \& {Smith}(2002)}]{Remillard02_ATEL88}
{Remillard}, R.~A. \& {Smith}, D. 2002, {Atel} 88

\bibitem[{{Revnivtsev} {et~al.}(2002){Revnivtsev}, {Gilfanov}, {Churazov}, \&
  {Sunyaev}}]{rgc+02}
{Revnivtsev}, M., {Gilfanov}, M., {Churazov}, E., \& {Sunyaev}, R. 2002, \aap,
  391, 1013

\bibitem[{{Schlegel} {et~al.}(1998){Schlegel}, {Finkbeiner}, \&
  {Davis}}]{sfd98}
{Schlegel}, D.~J., {Finkbeiner}, D.~P., \& {Davis}, M. 1998, \apj, 500, 525

\bibitem[{{Skrutskie} {et~al.}(2006){Skrutskie}, {Cutri}, {Stiening},
  {Weinberg}, {Schneider}, {Carpenter}, {Beichman}, {Capps}, {Chester},
  {Elias}, {Huchra}, {Liebert}, {Lonsdale}, {Monet}, {Price}, {Seitzer},
  {Jarrett}, {Kirkpatrick}, {Gizis}, {Howard}, {Evans}, {Fowler}, {Fullmer},
  {Hurt}, {Light}, {Kopan}, {Marsh}, {McCallon}, {Tam}, {Van Dyk}, \&
  {Wheelock}}]{scs+06}
{Skrutskie}, M.~F., {Cutri}, R.~M., {Stiening}, R., {Weinberg}, M.~D.,
  {Schneider}, S., {Carpenter}, J.~M., {Beichman}, C., {Capps}, R., {Chester},
  T., {Elias}, J., {Huchra}, J., {Liebert}, J., {Lonsdale}, C., {Monet}, D.~G.,
  {Price}, S., {Seitzer}, P., {Jarrett}, T., {Kirkpatrick}, J.~D., {Gizis},
  J.~E., {Howard}, E., {Evans}, T., {Fowler}, J., {Fullmer}, L., {Hurt}, R.,
  {Light}, R., {Kopan}, E.~L., {Marsh}, K.~A., {McCallon}, H.~L., {Tam}, R.,
  {Van Dyk}, S., \& {Wheelock}, S. 2006, \aj, 131, 1163

\bibitem[{{Stefanescu} {et~al.}(2007{\natexlab{a}}){Stefanescu}, {Slowikowska},
  {Kanbach}, {Duscha}, {Schrey}, {Steinle}, \&
  {Ioannou}}]{Stefanescu07_GCN6492}
{Stefanescu}, A., {Slowikowska}, A., {Kanbach}, G., {Duscha}, S., {Schrey}, F.,
  {Steinle}, H., \& {Ioannou}, Z. 2007{\natexlab{a}}, {GCN Circular} 6492

\bibitem[{{Stefanescu} {et~al.}(2007{\natexlab{b}}){Stefanescu}, {Slowikowska},
  {Kanbach}, {et~al.}}]{GCN.6508}
{Stefanescu}, A., {Slowikowska}, A., {Kanbach}, G., {et~al.}
  2007{\natexlab{b}}, {GCN Circular} 6508

\bibitem[{{Stefanescu} {et~al.}(2007{\natexlab{c}}){Stefanescu}, {Slowikowska},
  {Kanbach}, {et~al.}}]{GCN.6532}
---. 2007{\natexlab{c}}, {GCN Circular} 6532

\bibitem[{{Stern} {et~al.}(2001){Stern}, {Beloborodov}, \& {Poutanen}}]{sbp+01}
{Stern}, B.~E., {Beloborodov}, A.~M., \& {Poutanen}, J. 2001, \apj, 555, 829

\bibitem[{{Svensson}(1987)}]{s87}
{Svensson}, R. 1987, \mnras, 227, 403

\bibitem[{{Tanaka} \& {Shibazaki}(1996)}]{ts96}
{Tanaka}, Y. \& {Shibazaki}, N. 1996, \araa, 34, 607

\bibitem[{{Tueller} {et~al.}(2007){Tueller}, {Barbier}, {Barthelmy},
  {Cummings}, {Fenimore}, {Gehrels}, {Krimm}, {Markwardt}, {Pagani}, {Palmer},
  {Parsons}, {Sakamoto}, {Sato}, {Stamatikos}, \&
  {Ukwatta}}]{Tueller07_GCN6491}
{Tueller}, J., {Barbier}, L., {Barthelmy}, S.~D., {Cummings}, J., {Fenimore},
  E., {Gehrels}, N., {Krimm}, H., {Markwardt}, C., {Pagani}, C., {Palmer}, D.,
  {Parsons}, A., {Sakamoto}, T., {Sato}, G., {Stamatikos}, M., \& {Ukwatta}, T.
  2007, {GCN Circular} 6491

\bibitem[{{Uemura} {et~al.}(2004){Uemura}, {Kato}, {Ishioka}, {Imada},
  {Nogami}, {Monard}, {Cook}, {Stubbings}, {Kiyota}, {Nelson}, {Beninger},
  {Bolt}, \& {Heathcote}}]{uki+04}
{Uemura}, M., {Kato}, T., {Ishioka}, R., {Imada}, A., {Nogami}, D., {Monard},
  B., {Cook}, L.~M., {Stubbings}, R., {Kiyota}, S., {Nelson}, P., {Beninger},
  J.-Y., {Bolt}, G., \& {Heathcote}, B. 2004, \pasj, 56, 823

\bibitem[{{Uemura} {et~al.}(2002){Uemura}, {Kato}, {Watanabe}, {Stubbings},
  {Monard}, \& {Kawai}}]{ukw+02}
{Uemura}, M., {Kato}, T., {Watanabe}, T., {Stubbings}, R., {Monard}, B., \&
  {Kawai}, N. 2002, \pasj, 54, 95

\bibitem[{{Updike} {et~al.}(2007{\natexlab{a}}){Updike}, {Hartmann}, {Henson},
  {et~al.}}]{GCN.6507}
{Updike}, A.~C., {Hartmann}, D.~H., {Henson}, G., {et~al.} 2007{\natexlab{a}},
  {GCN Circular} 6507

\bibitem[{{Updike} {et~al.}(2007{\natexlab{b}}){Updike}, {Milne}, {Williams},
  {et~al.}}]{GCN.6536}
{Updike}, A.~C., {Milne}, P.~A., {Williams}, G.~G., {et~al.}
  2007{\natexlab{b}}, {GCN Circular} 6536

\bibitem[{{van Dam} {et~al.}(2006){van Dam}, {Bouchez}, {Le Mignant},
  {Johansson}, {Wizinowich}, {Campbell}, {Chin}, {Hartman}, {Lafon}, {Stomski},
  \& {Summers}}]{vbl+06}
{van Dam}, M.~A., {Bouchez}, A.~H., {Le Mignant}, D., {Johansson}, E.~M.,
  {Wizinowich}, P.~L., {Campbell}, R.~D., {Chin}, J.~C.~Y., {Hartman}, S.~K.,
  {Lafon}, R.~E., {Stomski}, Jr., P.~J., \& {Summers}, D.~M. 2006, \pasp, 118,
  310

\bibitem[{{Wilson} {et~al.}(2003){Wilson}, {Eikenberry}, {Henderson},
  {Hayward}, {Carson}, {Pirger}, {Barry}, {Brandl}, {Houck}, {Fitzgerald}, \&
  {Stolberg}}]{weh+03}
{Wilson}, J.~C., {Eikenberry}, S.~S., {Henderson}, C.~P., {Hayward}, T.~L.,
  {Carson}, J.~C., {Pirger}, B., {Barry}, D.~J., {Brandl}, B.~R., {Houck},
  J.~R., {Fitzgerald}, G.~J., \& {Stolberg}, T.~M. 2003, in Presented at the
  Society of Photo-Optical Instrumentation Engineers (SPIE) Conference, Vol.
  4841, Instrument Design and Performance for Optical/Infrared Ground-based
  Telescopes. Edited by Iye, Masanori; Moorwood, Alan F. M. Proceedings of the
  SPIE, Volume 4841, pp. 451-458 (2003)., ed. M.~{Iye} \& A.~F.~M. {Moorwood},
  451--458

\bibitem[{{Wizinowich} {et~al.}(2006){Wizinowich}, {Chin}, {Johansson},
  {Kellner}, {Lafon}, {Le Mignant}, {Neyman}, {Stomski}, {Summers}, {Sumner},
  \& {van Dam}}]{wcj+06}
{Wizinowich}, P.~L., {Chin}, J., {Johansson}, E., {Kellner}, S., {Lafon}, R.,
  {Le Mignant}, D., {Neyman}, C., {Stomski}, P., {Summers}, D., {Sumner}, R.,
  \& {van Dam}, M. 2006, in Presented at the Society of Photo-Optical
  Instrumentation Engineers (SPIE) Conference, Vol. 6272, Advances in Adaptive
  Optics II. Edited by Ellerbroek, Brent L.; Bonaccini Calia, Domenico.
  Proceedings of the SPIE, Volume 6272, pp. 627209 (2006).

\bibitem[{{Woods} {et~al.}(1999){Woods}, {Kouveliotou}, {van Paradijs},
  {Briggs}, {Wilson}, {Deal}, {Harmon}, {Fishman}, {Lewin}, \&
  {Kommers}}]{wkv+99}
{Woods}, P.~M., {Kouveliotou}, C., {van Paradijs}, J., {Briggs}, M.~S.,
  {Wilson}, C.~A., {Deal}, K., {Harmon}, B.~A., {Fishman}, G.~J., {Lewin},
  W.~H.~G., \& {Kommers}, J. 1999, \apj, 517, 431

\bibitem[{{Woods} \& {Thompson}(2006)}]{wt06}
{Woods}, P.~M. \& {Thompson}, C. 2006, {Soft gamma repeaters and anomalous
  X-ray pulsars: magnetar candidates} (Compact stellar X-ray sources), 547--586

\bibitem[{{Yoshida} {et~al.}(2007){Yoshida}, {Yanagisawa}, {Shimizu},
  {et~al.}}]{GCN.6512}
{Yoshida}, M., {Yanagisawa}, K., {Shimizu}, Y., {et~al.} 2007, {GCN Circular}
  6512

\bibitem[{{Zhang} {et~al.}(2006){Zhang}, {Fan}, {Dyks}, {Kobayashi},
  {M{\'e}sz{\'a}ros}, {Burrows}, {Nousek}, \& {Gehrels}}]{zfd+06}
{Zhang}, B., {Fan}, Y.~Z., {Dyks}, J., {Kobayashi}, S., {M{\'e}sz{\'a}ros}, P.,
  {Burrows}, D.~N., {Nousek}, J.~A., \& {Gehrels}, N. 2006, \apj, 642, 354

\end{thebibliography}

%\include{tab1}

\begin{deluxetable}{llllll}
  \tabletypesize{\footnotesize}
  \tablecaption{Optical Observations of \event\ with Palomar 60-inch telescope}
  \tablecolumns{6}
  \tablewidth{0pc}
 \tablehead{\colhead{Epoch} & \colhead{Facility} &
             \colhead{Filter} & \colhead{Time Since Burst} &
             \colhead{Exposure} & \colhead{Magnitude} \\
             \colhead{(2007 UT)} & & & \colhead{(hr)} & \colhead{(s)} &
            }
  \startdata
June 12.2416 & P60 & i' & 32.92 & 30.0 $\times$ 1 &    19.1 $\pm$ 0.14 \\
June 12.2422 & P60 & i' & 32.94 & 30.0 $\times$ 1 & $>$19.8 \\
June 12.2433 & P60 & i' & 32.96 & 30.0 $\times$ 3 & $>$19.8 \\
June 12.2444 & P60 & i' & 32.99 & 30.0 $\times$ 5 &    19.7 $\pm$ 0.16 \\
June 12.2450 & P60 & i' & 33.00 & 30.0 $\times$ 5 &    19.6 $\pm$ 0.15 \\
June 12.2456 & P60 & i' & 33.02 & 30.0 $\times$ 5 &    19.6 $\pm$ 0.15 \\
June 12.2462 & P60 & i' & 33.03 & 30.0 $\times$ 5 &    19.7 $\pm$ 0.15 \\
June 12.2467 & P60 & i' & 33.05 & 30.0 $\times$ 5 & $>$19.8 \\
June 12.2473 & P60 & i' & 33.06 & 30.0 $\times$ 5 & $>$19.8 \\
June 12.2479 & P60 & i' & 33.07 & 30.0 $\times$ 5 & $>$19.8 \\
June 12.2484 & P60 & i' & 33.09 & 30.0 $\times$ 1 & $>$19.8 \\
June 12.2484 & P60 & i' & 33.09 & 30.0 $\times$ 5 & $>$19.8 \\
June 12.2490 & P60 & i' & 33.10 & 30.0 $\times$ 1 &    18.9 $\pm$ 0.13 \\
June 12.2496 & P60 & i' & 33.11 & 30.0 $\times$ 1 &    18.7 $\pm$ 0.12 \\
June 12.2501 & P60 & i' & 33.13 & 30.0 $\times$ 1 & $>$19.8 \\
June 12.2513 & P60 & i' & 33.15 & 30.0 $\times$ 5 & $>$19.8 \\
June 12.2518 & P60 & i' & 33.17 & 30.0 $\times$ 3 &    19.8 $\pm$ 0.18 \\
June 12.2518 & P60 & i' & 33.17 & 30.0 $\times$ 5 &    19.7 $\pm$ 0.17 \\
June 12.2524 & P60 & i' & 33.18 & 30.0 $\times$ 3 &    19.7 $\pm$ 0.16 \\
June 12.2529 & P60 & i' & 33.19 & 30.0 $\times$ 3 &    19.7 $\pm$ 0.16 \\
June 12.2541 & P60 & i' & 33.22 & 30.0 $\times$ 5 & $>$19.9 \\
June 12.2547 & P60 & i' & 33.23 & 30.0 $\times$ 1 & $>$19.8 \\
June 12.2547 & P60 & i' & 33.23 & 30.0 $\times$ 5 & $>$19.9 \\
June 12.2552 & P60 & i' & 33.25 & 30.0 $\times$ 1 &    18.5 $\pm$ 0.14 \\
June 12.2558 & P60 & i' & 33.26 & 30.0 $\times$ 1 & $>$19.8 \\
June 12.2569 & P60 & i' & 33.29 & 30.0 $\times$ 1 & $>$19.9 \\
June 12.2569 & P60 & i' & 33.29 & 30.0 $\times$ 3 & $>$19.9 \\
June 12.2569 & P60 & i' & 33.29 & 30.0 $\times$ 5 & $>$19.9 \\
June 12.2575 & P60 & i' & 33.30 & 30.0 $\times$ 1 &    19.2 $\pm$ 0.15 \\
June 12.2581 & P60 & i' & 33.32 & 30.0 $\times$ 1 &    19.8 $\pm$ 0.17 \\
June 12.2598 & P60 & i' & 33.36 & 30.0 $\times$ 5 &    19.9 $\pm$ 0.16 \\
June 12.2604 & P60 & i' & 33.37 & 30.0 $\times$ 5 & $>$19.9 \\
June 12.2610 & P60 & i' & 33.39 & 30.0 $\times$ 5 & $>$19.9 \\
June 12.2615 & P60 & i' & 33.40 & 30.0 $\times$ 5 & $>$19.9 \\
June 12.2621 & P60 & i' & 33.41 & 30.0 $\times$ 5 & $>$20.0 \\
June 12.2626 & P60 & i' & 33.43 & 30.0 $\times$ 5 & $>$20.0 \\
June 12.2632 & P60 & i' & 33.44 & 30.0 $\times$ 3 &    19.9 $\pm$ 0.16 \\
June 12.2632 & P60 & i' & 33.44 & 30.0 $\times$ 5 & $>$20.0 \\
June 12.2638 & P60 & i' & 33.45 & 30.0 $\times$ 3 &    19.7 $\pm$ 0.16 \\
June 12.2644 & P60 & i' & 33.47 & 30.0 $\times$ 3 &    19.6 $\pm$ 0.15 \\
June 12.2649 & P60 & i' & 33.48 & 30.0 $\times$ 3 &    19.8 $\pm$ 0.16 \\
June 12.2655 & P60 & i' & 33.50 & 30.0 $\times$ 3 &    19.9 $\pm$ 0.17 \\
June 12.2666 & P60 & i' & 33.52 & 30.0 $\times$ 3 & $>$20.0 \\
June 12.2666 & P60 & i' & 33.52 & 30.0 $\times$ 5 &    19.8 $\pm$ 0.16 \\
June 12.2678 & P60 & i' & 33.55 & 30.0 $\times$ 1 &    19.4 $\pm$ 0.14 \\
June 12.2678 & P60 & i' & 33.55 & 30.0 $\times$ 3 & $>$20.0 \\
June 12.2678 & P60 & i' & 33.55 & 30.0 $\times$ 5 &    19.9 $\pm$ 0.16 \\
June 12.2689 & P60 & i' & 33.58 & 30.0 $\times$ 3 &    19.9 $\pm$ 0.17 \\
June 12.3132 & P60 & i' & 34.64 & 30.0 $\times$ 3 &    20.2 $\pm$ 0.19 \\
June 12.3225 & P60 & i' & 34.86 & 30.0 $\times$ 3 & $>$20.4 \\
June 12.3230 & P60 & i' & 34.88 & 30.0 $\times$ 3 &    19.9 $\pm$ 0.16 \\
June 12.3236 & P60 & i' & 34.89 & 30.0 $\times$ 3 &    20.1 $\pm$ 0.16 \\
June 12.3242 & P60 & i' & 34.90 & 30.0 $\times$ 3 &    20.3 $\pm$ 0.19 \\
June 12.3247 & P60 & i' & 34.92 & 30.0 $\times$ 3 &    20.3 $\pm$ 0.19 \\
June 12.3253 & P60 & i' & 34.93 & 30.0 $\times$ 1 & $>$20.3 \\
June 12.3253 & P60 & i' & 34.93 & 30.0 $\times$ 3 &    20.2 $\pm$ 0.19 \\
June 12.3259 & P60 & i' & 34.94 & 30.0 $\times$ 1 &    19.1 $\pm$ 0.13 \\
June 12.3265 & P60 & i' & 34.96 & 30.0 $\times$ 1 &    19.4 $\pm$ 0.14 \\
June 12.3270 & P60 & i' & 34.97 & 30.0 $\times$ 1 & $>$20.3 \\
June 12.3288 & P60 & i' & 35.01 & 30.0 $\times$ 5 & $>$20.3 \\
June 12.3293 & P60 & i' & 35.03 & 30.0 $\times$ 5 & $>$20.3 \\
June 12.3299 & P60 & i' & 35.04 & 30.0 $\times$ 5 & $>$20.3 \\
June 12.3305 & P60 & i' & 35.05 & 30.0 $\times$ 5 & $>$20.4 \\
June 12.3311 & P60 & i' & 35.07 & 30.0 $\times$ 5 &    20.3 $\pm$ 0.17 \\
June 12.3316 & P60 & i' & 35.08 & 30.0 $\times$ 5 &    20.2 $\pm$ 0.18 \\
June 12.3322 & P60 & i' & 35.10 & 30.0 $\times$ 5 & $>$20.4 \\
June 12.3328 & P60 & i' & 35.11 & 30.0 $\times$ 5 & $>$20.5 \\
June 12.3334 & P60 & i' & 35.12 & 30.0 $\times$ 5 & $>$20.5 \\
June 12.3339 & P60 & i' & 35.14 & 30.0 $\times$ 3 &    20.4 $\pm$ 0.17 \\
June 12.3339 & P60 & i' & 35.14 & 30.0 $\times$ 5 &    20.2 $\pm$ 0.17 \\
June 12.3345 & P60 & i' & 35.15 & 30.0 $\times$ 3 &    20.3 $\pm$ 0.18 \\
June 12.3351 & P60 & i' & 35.17 & 30.0 $\times$ 3 &    20.1 $\pm$ 0.16 \\
June 12.3357 & P60 & i' & 35.18 & 30.0 $\times$ 3 &    20.3 $\pm$ 0.17 \\
June 12.3363 & P60 & i' & 35.19 & 30.0 $\times$ 1 &    20.2 $\pm$ 0.20 \\
June 12.3363 & P60 & i' & 35.19 & 30.0 $\times$ 3 &    20.1 $\pm$ 0.17 \\
June 12.3369 & P60 & i' & 35.21 & 30.0 $\times$ 1 &    19.7 $\pm$ 0.17 \\
June 12.3380 & P60 & i' & 35.24 & 30.0 $\times$ 3 &    20.0 $\pm$ 0.17 \\
June 12.3386 & P60 & i' & 35.25 & 30.0 $\times$ 3 &    20.3 $\pm$ 0.17 \\
June 12.3392 & P60 & i' & 35.26 & 30.0 $\times$ 1 & $>$20.3 \\
June 12.3392 & P60 & i' & 35.26 & 30.0 $\times$ 3 &    20.3 $\pm$ 0.19 \\
June 12.3397 & P60 & i' & 35.28 & 30.0 $\times$ 1 &    19.1 $\pm$ 0.13 \\
June 12.3403 & P60 & i' & 35.29 & 30.0 $\times$ 1 & $>$20.3 \\
June 12.3409 & P60 & i' & 35.30 & 30.0 $\times$ 1 & $>$20.3 \\
June 12.3420 & P60 & i' & 35.33 & 30.0 $\times$ 1 &    20.0 $\pm$ 0.16 \\
June 12.3420 & P60 & i' & 35.33 & 30.0 $\times$ 3 &    20.0 $\pm$ 0.15 \\
June 12.3432 & P60 & i' & 35.36 & 30.0 $\times$ 3 &    20.1 $\pm$ 0.17 \\
June 12.3443 & P60 & i' & 35.39 & 30.0 $\times$ 5 & $>$20.4 \\
June 12.3449 & P60 & i' & 35.40 & 30.0 $\times$ 5 & $>$20.4 \\
June 12.3455 & P60 & i' & 35.42 & 30.0 $\times$ 3 &    20.2 $\pm$ 0.18 \\
June 12.3455 & P60 & i' & 35.42 & 30.0 $\times$ 5 &    20.2 $\pm$ 0.18 \\
June 12.3466 & P60 & i' & 35.44 & 30.0 $\times$ 5 &    20.2 $\pm$ 0.18 \\
June 12.3472 & P60 & i' & 35.46 & 30.0 $\times$ 5 & $>$20.3 \\
June 12.3478 & P60 & i' & 35.47 & 30.0 $\times$ 5 & $>$20.3 \\
June 12.3484 & P60 & i' & 35.48 & 30.0 $\times$ 5 & $>$20.3 \\
June 12.3490 & P60 & i' & 35.50 & 30.0 $\times$ 5 & $>$20.3 \\
June 12.3495 & P60 & i' & 35.51 & 30.0 $\times$ 1 & $>$20.3 \\
June 12.3495 & P60 & i' & 35.51 & 30.0 $\times$ 5 & $>$20.4 \\
June 12.3501 & P60 & i' & 35.53 & 30.0 $\times$ 1 & $>$20.3 \\
June 12.3507 & P60 & i' & 35.54 & 30.0 $\times$ 1 & $>$20.3 \\
June 12.3513 & P60 & i' & 35.55 & 30.0 $\times$ 1 & $>$20.2 \\
June 12.3524 & P60 & i' & 35.58 & 30.0 $\times$ 5 & $>$20.4 \\
June 12.3530 & P60 & i' & 35.60 & 30.0 $\times$ 5 & $>$20.3 \\
June 12.3536 & P60 & i' & 35.61 & 30.0 $\times$ 5 & $>$20.4 \\
June 12.3542 & P60 & i' & 35.62 & 30.0 $\times$ 5 & $>$20.4 \\
June 12.3547 & P60 & i' & 35.64 & 30.0 $\times$ 5 & $>$20.4 \\
June 12.3553 & P60 & i' & 35.65 & 30.0 $\times$ 5 & $>$20.4 \\
June 12.3559 & P60 & i' & 35.66 & 30.0 $\times$ 5 &    20.4 $\pm$ 0.18 \\
June 12.3565 & P60 & i' & 35.68 & 30.0 $\times$ 5 &    20.3 $\pm$ 0.18 \\
June 12.3571 & P60 & i' & 35.69 & 30.0 $\times$ 5 &    20.3 $\pm$ 0.18 \\
June 12.3576 & P60 & i' & 35.71 & 30.0 $\times$ 5 & $>$20.3 \\
June 12.3582 & P60 & i' & 35.72 & 30.0 $\times$ 5 & $>$20.3 \\
June 12.3588 & P60 & i' & 35.73 & 30.0 $\times$ 3 &    20.3 $\pm$ 0.17 \\
June 12.3588 & P60 & i' & 35.73 & 30.0 $\times$ 5 &    20.3 $\pm$ 0.17 \\
June 12.3594 & P60 & i' & 35.75 & 30.0 $\times$ 3 &    20.2 $\pm$ 0.18 \\
June 12.3605 & P60 & i' & 35.78 & 30.0 $\times$ 5 &    20.2 $\pm$ 0.18 \\
June 12.3611 & P60 & i' & 35.79 & 30.0 $\times$ 5 &    20.3 $\pm$ 0.18 \\
June 12.3617 & P60 & i' & 35.80 & 30.0 $\times$ 5 & $>$20.4 \\
June 12.3623 & P60 & i' & 35.82 & 30.0 $\times$ 5 &    20.3 $\pm$ 0.18 \\
June 12.3628 & P60 & i' & 35.83 & 30.0 $\times$ 3 &    20.3 $\pm$ 0.20 \\
June 12.3628 & P60 & i' & 35.83 & 30.0 $\times$ 5 &    20.1 $\pm$ 0.18 \\
June 12.3635 & P60 & i' & 35.85 & 30.0 $\times$ 3 &    20.1 $\pm$ 0.18 \\
June 12.3641 & P60 & i' & 35.86 & 30.0 $\times$ 3 &    20.0 $\pm$ 0.17 \\
June 12.3646 & P60 & i' & 35.87 & 30.0 $\times$ 3 &    20.2 $\pm$ 0.18 \\
June 12.3652 & P60 & i' & 35.89 & 30.0 $\times$ 3 & $>$20.4 \\
June 12.3658 & P60 & i' & 35.90 & 30.0 $\times$ 3 &    20.2 $\pm$ 0.17 \\
June 12.3670 & P60 & i' & 35.93 & 30.0 $\times$ 1 & $>$20.4 \\
June 12.3670 & P60 & i' & 35.93 & 30.0 $\times$ 3 &    19.9 $\pm$ 0.16 \\
June 12.3676 & P60 & i' & 35.95 & 30.0 $\times$ 1 &    20.0 $\pm$ 0.16 \\
June 12.3682 & P60 & i' & 35.96 & 30.0 $\times$ 1 &    20.2 $\pm$ 0.19 \\
June 12.3694 & P60 & i' & 35.99 & 30.0 $\times$ 3 &    20.3 $\pm$ 0.18 \\
June 12.3700 & P60 & i' & 36.00 & 30.0 $\times$ 3 &    20.2 $\pm$ 0.16 \\
June 12.3705 & P60 & i' & 36.02 & 30.0 $\times$ 3 &    20.0 $\pm$ 0.16 \\
June 12.3711 & P60 & i' & 36.03 & 30.0 $\times$ 3 &    20.0 $\pm$ 0.16 \\
June 12.3717 & P60 & i' & 36.04 & 30.0 $\times$ 1 & $>$20.4 \\
June 12.3717 & P60 & i' & 36.04 & 30.0 $\times$ 3 & $>$20.4 \\
June 12.3729 & P60 & i' & 36.07 & 30.0 $\times$ 3 &    20.2 $\pm$ 0.19 \\
June 12.3735 & P60 & i' & 36.09 & 30.0 $\times$ 3 &    20.2 $\pm$ 0.17 \\
June 12.3740 & P60 & i' & 36.10 & 30.0 $\times$ 3 & $>$20.4 \\
June 12.3746 & P60 & i' & 36.11 & 30.0 $\times$ 3 &    20.3 $\pm$ 0.17 \\
June 12.3752 & P60 & i' & 36.13 & 30.0 $\times$ 1 & $>$20.4 \\
June 12.3752 & P60 & i' & 36.13 & 30.0 $\times$ 3 &    20.2 $\pm$ 0.18 \\
June 12.3764 & P60 & i' & 36.16 & 30.0 $\times$ 3 &    20.3 $\pm$ 0.18 \\
June 12.3776 & P60 & i' & 36.18 & 30.0 $\times$ 5 &    20.3 $\pm$ 0.17 \\
June 12.3782 & P60 & i' & 36.20 & 30.0 $\times$ 1 & $>$20.4 \\
June 12.3782 & P60 & i' & 36.20 & 30.0 $\times$ 5 & $>$20.4 \\
June 12.3787 & P60 & i' & 36.21 & 30.0 $\times$ 1 &    16.7 $\pm$ 0.12 \\
June 12.3793 & P60 & i' & 36.23 & 30.0 $\times$ 1 &    19.2 $\pm$ 0.12 \\
June 12.3799 & P60 & i' & 36.24 & 30.0 $\times$ 1 &    19.0 $\pm$ 0.13 \\
June 12.3805 & P60 & i' & 36.26 & 30.0 $\times$ 1 & $>$20.4 \\
June 12.3817 & P60 & i' & 36.28 & 30.0 $\times$ 3 &    20.0 $\pm$ 0.16 \\
June 12.3817 & P60 & i' & 36.28 & 30.0 $\times$ 5 &    20.4 $\pm$ 0.17 \\
June 12.3823 & P60 & i' & 36.30 & 30.0 $\times$ 3 &    20.3 $\pm$ 0.18 \\
June 12.3829 & P60 & i' & 36.31 & 30.0 $\times$ 3 &    20.3 $\pm$ 0.18 \\
June 12.3840 & P60 & i' & 36.34 & 30.0 $\times$ 1 & $>$20.4 \\
June 12.3840 & P60 & i' & 36.34 & 30.0 $\times$ 3 & $>$20.4 \\
June 12.3840 & P60 & i' & 36.34 & 30.0 $\times$ 5 &    20.3 $\pm$ 0.18 \\
June 12.3846 & P60 & i' & 36.35 & 30.0 $\times$ 1 &    18.5 $\pm$ 0.13 \\
June 12.3852 & P60 & i' & 36.37 & 30.0 $\times$ 1 &    19.9 $\pm$ 0.16 \\
June 12.3858 & P60 & i' & 36.38 & 30.0 $\times$ 1 & $>$20.5 \\
June 12.3869 & P60 & i' & 36.41 & 30.0 $\times$ 3 &    20.0 $\pm$ 0.17 \\
June 12.3875 & P60 & i' & 36.42 & 30.0 $\times$ 3 &    20.1 $\pm$ 0.17 \\
June 12.3881 & P60 & i' & 36.44 & 30.0 $\times$ 1 & $>$20.5 \\
June 12.3881 & P60 & i' & 36.44 & 30.0 $\times$ 3 &    20.2 $\pm$ 0.19 \\
June 12.3887 & P60 & i' & 36.45 & 30.0 $\times$ 1 &    17.5 $\pm$ 0.11 \\
June 12.3893 & P60 & i' & 36.47 & 30.0 $\times$ 1 &    19.3 $\pm$ 0.14 \\
June 12.3899 & P60 & i' & 36.48 & 30.0 $\times$ 1 & $>$20.4 \\
June 12.3917 & P60 & i' & 36.52 & 30.0 $\times$ 5 &    20.3 $\pm$ 0.17 \\
June 12.3923 & P60 & i' & 36.54 & 30.0 $\times$ 3 &    20.3 $\pm$ 0.18 \\
June 12.3923 & P60 & i' & 36.54 & 30.0 $\times$ 5 &    20.3 $\pm$ 0.17 \\
June 12.3935 & P60 & i' & 36.57 & 30.0 $\times$ 5 &    20.4 $\pm$ 0.18 \\
June 12.3940 & P60 & i' & 36.58 & 30.0 $\times$ 5 &    20.4 $\pm$ 0.19 \\
June 12.3946 & P60 & i' & 36.59 & 30.0 $\times$ 1 & $>$20.4 \\
June 12.3946 & P60 & i' & 36.59 & 30.0 $\times$ 5 &    20.4 $\pm$ 0.19 \\
June 12.3952 & P60 & i' & 36.61 & 30.0 $\times$ 1 &    19.6 $\pm$ 0.15 \\
June 12.3958 & P60 & i' & 36.62 & 30.0 $\times$ 1 &    20.0 $\pm$ 0.15 \\
June 12.3970 & P60 & i' & 36.65 & 30.0 $\times$ 3 &    20.3 $\pm$ 0.19 \\
June 12.3970 & P60 & i' & 36.65 & 30.0 $\times$ 5 &    20.4 $\pm$ 0.19 \\
June 12.3976 & P60 & i' & 36.67 & 30.0 $\times$ 3 &    20.3 $\pm$ 0.19 \\
June 12.3982 & P60 & i' & 36.68 & 30.0 $\times$ 1 & $>$20.4 \\
June 12.3982 & P60 & i' & 36.68 & 30.0 $\times$ 3 &    20.4 $\pm$ 0.18 \\
June 12.3988 & P60 & i' & 36.69 & 30.0 $\times$ 1 &    18.7 $\pm$ 0.13 \\
June 12.3993 & P60 & i' & 36.71 & 30.0 $\times$ 1 & $>$20.5 \\
June 12.4005 & P60 & i' & 36.74 & 30.0 $\times$ 3 &    20.2 $\pm$ 0.18 \\
June 12.4011 & P60 & i' & 36.75 & 30.0 $\times$ 3 &    20.3 $\pm$ 0.19 \\
June 12.4017 & P60 & i' & 36.76 & 30.0 $\times$ 3 &    20.4 $\pm$ 0.18 \\
June 12.4029 & P60 & i' & 36.79 & 30.0 $\times$ 5 &    20.5 $\pm$ 0.19 \\
June 12.4035 & P60 & i' & 36.81 & 30.0 $\times$ 5 &    20.4 $\pm$ 0.19 \\
June 12.4041 & P60 & i' & 36.82 & 30.0 $\times$ 3 &    20.3 $\pm$ 0.18 \\
June 12.4041 & P60 & i' & 36.82 & 30.0 $\times$ 5 &    20.3 $\pm$ 0.18 \\
June 12.4047 & P60 & i' & 36.84 & 30.0 $\times$ 3 &    20.1 $\pm$ 0.18 \\
June 12.4052 & P60 & i' & 36.85 & 30.0 $\times$ 1 & $>$20.5 \\
June 12.4052 & P60 & i' & 36.85 & 30.0 $\times$ 3 & $>$20.5 \\
June 12.4058 & P60 & i' & 36.86 & 30.0 $\times$ 1 & $>$20.6 \\
June 12.4064 & P60 & i' & 36.88 & 30.0 $\times$ 1 & $>$20.6 \\
June 12.4070 & P60 & i' & 36.89 & 30.0 $\times$ 1 & $>$20.6 \\
June 12.4076 & P60 & i' & 36.91 & 30.0 $\times$ 1 & $>$20.6 \\
June 12.4082 & P60 & i' & 36.92 & 30.0 $\times$ 1 & $>$20.6 \\
June 12.4088 & P60 & i' & 36.93 & 30.0 $\times$ 1 & $>$20.7 \\
June 12.4094 & P60 & i' & 36.95 & 30.0 $\times$ 1 & $>$20.7 \\
June 12.4118 & P60 & i' & 37.01 & 30.0 $\times$ 5 &    20.5 $\pm$ 0.18 \\
June 12.4124 & P60 & i' & 37.02 & 30.0 $\times$ 5 &    20.5 $\pm$ 0.18 \\
June 12.4130 & P60 & i' & 37.03 & 30.0 $\times$ 5 &    20.7 $\pm$ 0.20 \\
June 12.4136 & P60 & i' & 37.05 & 30.0 $\times$ 3 &    20.5 $\pm$ 0.17 \\
June 12.4136 & P60 & i' & 37.05 & 30.0 $\times$ 5 &    20.7 $\pm$ 0.20 \\
June 12.4142 & P60 & i' & 37.06 & 30.0 $\times$ 3 &    20.6 $\pm$ 0.19 \\
June 12.4153 & P60 & i' & 37.09 & 30.0 $\times$ 5 &    20.5 $\pm$ 0.19 \\
June 12.4159 & P60 & i' & 37.11 & 30.0 $\times$ 5 & $>$20.7 \\
June 12.4165 & P60 & i' & 37.12 & 30.0 $\times$ 3 &    20.6 $\pm$ 0.20 \\
June 12.4165 & P60 & i' & 37.12 & 30.0 $\times$ 5 &    20.7 $\pm$ 0.21 \\
June 12.4171 & P60 & i' & 37.13 & 30.0 $\times$ 1 & $>$20.7 \\
June 12.4171 & P60 & i' & 37.13 & 30.0 $\times$ 3 &    20.5 $\pm$ 0.18 \\
June 12.4177 & P60 & i' & 37.15 & 30.0 $\times$ 1 & $>$20.7 \\
June 12.4183 & P60 & i' & 37.16 & 30.0 $\times$ 1 & $>$20.7 \\
June 12.4195 & P60 & i' & 37.19 & 30.0 $\times$ 1 & $>$20.6 \\
June 12.4195 & P60 & i' & 37.19 & 30.0 $\times$ 3 &    20.6 $\pm$ 0.19 \\
June 12.4201 & P60 & i' & 37.21 & 30.0 $\times$ 1 &    19.6 $\pm$ 0.15 \\
June 12.4207 & P60 & i' & 37.22 & 30.0 $\times$ 1 &    20.4 $\pm$ 0.20 \\
June 12.4219 & P60 & i' & 37.25 & 30.0 $\times$ 1 &    20.0 $\pm$ 0.19 \\
June 12.4219 & P60 & i' & 37.25 & 30.0 $\times$ 3 &    20.0 $\pm$ 0.17 \\
June 12.4225 & P60 & i' & 37.26 & 30.0 $\times$ 1 &    19.3 $\pm$ 0.14 \\
June 12.4231 & P60 & i' & 37.28 & 30.0 $\times$ 1 &    20.4 $\pm$ 0.18 \\
June 12.4243 & P60 & i' & 37.31 & 30.0 $\times$ 3 &    20.5 $\pm$ 0.18 \\
June 12.4249 & P60 & i' & 37.32 & 30.0 $\times$ 3 & $>$20.7 \\
June 12.4255 & P60 & i' & 37.33 & 30.0 $\times$ 1 & $>$20.6 \\
June 12.4255 & P60 & i' & 37.33 & 30.0 $\times$ 3 &    20.6 $\pm$ 0.21 \\
June 12.4261 & P60 & i' & 37.35 & 30.0 $\times$ 1 & $>$20.6 \\
June 12.4267 & P60 & i' & 37.36 & 30.0 $\times$ 1 & $>$20.6 \\
June 12.4273 & P60 & i' & 37.38 & 30.0 $\times$ 1 & $>$20.6 \\
June 12.4279 & P60 & i' & 37.39 & 30.0 $\times$ 1 &    19.6 $\pm$ 0.15 \\
June 12.4285 & P60 & i' & 37.41 & 30.0 $\times$ 1 & $>$20.6 \\
June 12.4291 & P60 & i' & 37.42 & 30.0 $\times$ 1 & $>$20.6 \\
June 12.4297 & P60 & i' & 37.44 & 30.0 $\times$ 1 &    19.7 $\pm$ 0.14 \\
June 12.4303 & P60 & i' & 37.45 & 30.0 $\times$ 1 &    19.8 $\pm$ 0.16 \\
June 12.4309 & P60 & i' & 37.46 & 30.0 $\times$ 1 & $>$20.6 \\
June 12.4321 & P60 & i' & 37.49 & 30.0 $\times$ 3 &    20.4 $\pm$ 0.19 \\
June 12.4327 & P60 & i' & 37.51 & 30.0 $\times$ 1 & $>$20.5 \\
June 12.4327 & P60 & i' & 37.51 & 30.0 $\times$ 3 & $>$20.6 \\
June 12.4333 & P60 & i' & 37.52 & 30.0 $\times$ 1 &    19.6 $\pm$ 0.13 \\
June 12.4339 & P60 & i' & 37.54 & 30.0 $\times$ 1 &    19.8 $\pm$ 0.14 \\
June 12.4345 & P60 & i' & 37.55 & 30.0 $\times$ 1 & $>$20.5 \\
June 12.4357 & P60 & i' & 37.58 & 30.0 $\times$ 3 &    20.4 $\pm$ 0.19 \\
June 12.4363 & P60 & i' & 37.59 & 30.0 $\times$ 3 &    20.6 $\pm$ 0.20 \\
June 12.4368 & P60 & i' & 37.61 & 30.0 $\times$ 1 & $>$20.6 \\
June 12.4368 & P60 & i' & 37.61 & 30.0 $\times$ 3 &    20.5 $\pm$ 0.19 \\
June 12.4374 & P60 & i' & 37.62 & 30.0 $\times$ 1 &    18.5 $\pm$ 0.13 \\
June 12.4540 & P60 & i' & 38.02 & 30.0 $\times$ 1 &    20.1 $\pm$ 0.18 \\
June 12.4552 & P60 & i' & 38.05 & 30.0 $\times$ 3 &    20.4 $\pm$ 0.18 \\
June 12.4557 & P60 & i' & 38.06 & 30.0 $\times$ 3 &    20.4 $\pm$ 0.18 \\
June 12.4563 & P60 & i' & 38.08 & 30.0 $\times$ 1 &    20.2 $\pm$ 0.19 \\
June 12.4563 & P60 & i' & 38.08 & 30.0 $\times$ 3 &    20.2 $\pm$ 0.17 \\
June 12.4570 & P60 & i' & 38.09 & 30.0 $\times$ 1 &    19.9 $\pm$ 0.18 \\
June 12.4576 & P60 & i' & 38.10 & 30.0 $\times$ 1 &    20.4 $\pm$ 0.18 \\
June 12.4582 & P60 & i' & 38.12 & 30.0 $\times$ 1 &    20.1 $\pm$ 0.20 \\
June 12.4588 & P60 & i' & 38.13 & 30.0 $\times$ 1 & $>$20.4 \\
June 12.4594 & P60 & i' & 38.15 & 30.0 $\times$ 1 &    19.9 $\pm$ 0.17 \\
June 12.4600 & P60 & i' & 38.16 & 30.0 $\times$ 1 &    19.6 $\pm$ 0.15 \\
June 12.4606 & P60 & i' & 38.18 & 30.0 $\times$ 1 & $>$20.4 \\
June 12.4618 & P60 & i' & 38.21 & 30.0 $\times$ 3 &    20.0 $\pm$ 0.17 \\
June 12.4624 & P60 & i' & 38.22 & 30.0 $\times$ 3 &    20.2 $\pm$ 0.17 \\
June 12.4630 & P60 & i' & 38.24 & 30.0 $\times$ 3 &    20.4 $\pm$ 0.18 \\
June 12.4636 & P60 & i' & 38.25 & 30.0 $\times$ 3 &    20.2 $\pm$ 0.16 \\
June 12.4642 & P60 & i' & 38.26 & 30.0 $\times$ 1 & $>$20.4 \\
June 12.4642 & P60 & i' & 38.26 & 30.0 $\times$ 3 &    20.0 $\pm$ 0.15 \\
June 12.4648 & P60 & i' & 38.28 & 30.0 $\times$ 1 &    20.4 $\pm$ 0.20 \\
June 12.4654 & P60 & i' & 38.29 & 30.0 $\times$ 1 &    19.0 $\pm$ 0.13 \\
June 12.4660 & P60 & i' & 38.31 & 30.0 $\times$ 1 &    20.1 $\pm$ 0.16 \\
June 12.4666 & P60 & i' & 38.32 & 30.0 $\times$ 1 &    20.3 $\pm$ 0.18 \\
June 12.4679 & P60 & i' & 38.35 & 30.0 $\times$ 3 &    20.1 $\pm$ 0.17 \\
June 12.4685 & P60 & i' & 38.37 & 30.0 $\times$ 3 &    20.2 $\pm$ 0.18 \\
June 12.4691 & P60 & i' & 38.38 & 30.0 $\times$ 1 & $>$20.4 \\
June 12.4691 & P60 & i' & 38.38 & 30.0 $\times$ 3 &    20.1 $\pm$ 0.17 \\
June 12.4703 & P60 & i' & 38.41 & 30.0 $\times$ 3 &    20.0 $\pm$ 0.16 \\
June 12.4709 & P60 & i' & 38.43 & 30.0 $\times$ 3 &    20.0 $\pm$ 0.17 \\
June 12.4715 & P60 & i' & 38.44 & 30.0 $\times$ 3 & $>$20.4 \\
June 12.4721 & P60 & i' & 38.45 & 30.0 $\times$ 3 &    20.2 $\pm$ 0.18 \\
June 12.4727 & P60 & i' & 38.47 & 30.0 $\times$ 3 &    20.0 $\pm$ 0.15 \\
June 12.4733 & P60 & i' & 38.48 & 30.0 $\times$ 3 &    20.0 $\pm$ 0.15 \\
June 12.4739 & P60 & i' & 38.50 & 30.0 $\times$ 3 &    20.1 $\pm$ 0.16 \\
June 12.4748 & P60 & i' & 38.52 & 30.0 $\times$ 3 &    20.2 $\pm$ 0.17 \\
June 12.4754 & P60 & i' & 38.53 & 30.0 $\times$ 3 &    20.2 $\pm$ 0.18 \\
June 12.4766 & P60 & i' & 38.56 & 30.0 $\times$ 5 &    20.1 $\pm$ 0.18 \\
June 12.4772 & P60 & i' & 38.58 & 30.0 $\times$ 5 &    20.2 $\pm$ 0.18 \\
June 12.4778 & P60 & i' & 38.59 & 30.0 $\times$ 3 &    20.1 $\pm$ 0.17 \\
June 12.4778 & P60 & i' & 38.59 & 30.0 $\times$ 5 &    20.2 $\pm$ 0.17 \\
June 12.4784 & P60 & i' & 38.61 & 30.0 $\times$ 3 &    20.1 $\pm$ 0.17 \\
June 12.4790 & P60 & i' & 38.62 & 30.0 $\times$ 3 &    19.7 $\pm$ 0.15 \\
June 12.4796 & P60 & i' & 38.63 & 30.0 $\times$ 1 & $>$20.0 \\
June 12.4796 & P60 & i' & 38.63 & 30.0 $\times$ 3 & $>$20.1 \\
June 12.4816 & P60 & i' & 38.68 & 30.0 $\times$ 5 & $>$19.8 \\
June 12.4822 & P60 & i' & 38.70 & 30.0 $\times$ 1 &    19.4 $\pm$ 0.15 \\
June 12.4822 & P60 & i' & 38.70 & 30.0 $\times$ 5 &    19.7 $\pm$ 0.16 \\
June 13.2392 & P60 & i' & 56.86 & 60.0 $\times$ 1 & $>$20.2 \\
June 13.2484 & P60 & i' & 57.08 & 60.0 $\times$ 5 & $>$20.0 \\
June 14.4137 & P60 & i' & 85.05 & 180. $\times$ 1 & $>$18.4 \\
June 15.2329 & P60 & i' & 104.7 & 180. $\times$ 1 & $>$18.3 \\
June 16.4076 & P60 & i' & 132.9 & 180. $\times$ 1 & $>$18.8 \\
June 17.2273 & P60 & i' & 152.5 & 180. $\times$ 1 & $>$18.3 \\
June 18.3801 & P60 & i' & 180.2 & 180. $\times$ 1 & $>$18.4 \\
June 19.3444 & P60 & i' & 203.3 & 180. $\times$ 1 & $>$18.5 \\
June 20.3703 & P60 & i' & 228.0 & 180. $\times$ 1 & $>$18.4 \\
June 20.3889 & P60 & i' & 228.4 & 180. $\times$ 1 & $>$18.2 \\
 \enddata
  \tablecomments{Zeropoint computed in the AB system.
        The UT epoch denotes the start of observations.
        Error quoted are 1$\sigma$ photometric and instrumental
        errors summed in quadrature. Note that the uncertainty in the 
        zeropoint estimate (relative to USNO-B stars) dominates but is  
        an overestimate for variability studies using relative magnitude. 
        Upper limits quoted are 3$\sigma$.  No
        correction has been made for the large line-of-sight extinction.}
\label{tab:p60lc}
\end{deluxetable}

\end{document}